\documentclass[twocolumn]{aastex63}
\usepackage{appendix}
\usepackage{longtable}
\usepackage{booktabs}
\shorttitle{Detection of NH$_2$OH in the ISM}
\shortauthors{Rivilla et al.}

\begin{document}

\title{Prebiotic precursors of the primordial RNA world in space: Detection of NH$_2$OH
}

\author[0000-0002-2887-5859]{V\'ictor M. Rivilla}
\affiliation{Centro de Astrobiolog\'ia (CSIC-INTA), Ctra. de Ajalvir Km. 4, Torrej\'on de Ardoz, 28850 Madrid, Spain}
\affiliation{INAF-Osservatorio Astrofisico di Arcetri, Largo Enrico Fermi 5, 50125, Florence, Italy}

\author{Jes\'us Mart\'in-Pintado}
\author{Izaskun Jim\'enez-Serra}
\affiliation{Centro de Astrobiolog\'ia (CSIC-INTA), Ctra. de Ajalvir Km. 4, Torrej\'on de Ardoz, 28850 Madrid, Spain}

\author{Sergio Mart\'in}
\affiliation{European Southern Observatory, Alonso de C\'ordova 3107, Vitacura 763 0355, Santiago, Chile}
\affiliation{Joint ALMA Observatory, Alonso de C\'ordova 3107, Vitacura 763 0355, Santiago, Chile}

\author{Lucas F. Rodr\'iguez-Almeida}
\affiliation{Centro de Astrobiolog\'ia (CSIC-INTA), Ctra. de Ajalvir Km. 4, Torrej\'on de Ardoz, 28850 Madrid, Spain}

\author{Miguel A. Requena-Torres}
\affiliation{University of Maryland, College Park, ND 20742-2421 (USA)}
\affiliation{Department of Physics, Astronomy and Geosciences, Towson University, Towson, MD 21252, USA}

\author{Fernando Rico-Villas}
\affiliation{Centro de Astrobiolog\'ia (CSIC-INTA), Ctra. de Ajalvir Km. 4, Torrej\'on de Ardoz, 28850 Madrid, Spain}

\author{Shaoshan Zeng}
\affiliation{Star and Planet Formation Laboratory, Cluster for Pioneering Research, RIKEN, 2-1 Hirosawa, Wako, Saitama, 351-0198, Japan}

\author{Carlos Briones}
\affiliation{Centro de Astrobiolog\'ia (CSIC-INTA), Ctra. de Ajalvir Km. 4, Torrej\'on de Ardoz, 28850 Madrid, Spain}



\begin{abstract}

One of the proposed scenarios for the origin of life is the {\it primordial RNA world}, which considers that RNA molecules were likely responsible for the storage of genetic information and the catalysis of biochemical reactions in primitive cells, before the advent of proteins and DNA. In the last decade, experiments in the field of prebiotic chemistry have shown that RNA nucleotides can be synthesized from relatively simple molecular precursors, most of which have been found in space. 
An important exception is hydroxylamine, NH$_2$OH, which,  despite several observational attempts, it has not been detected in space yet.
Here we present the first detection of NH$_2$OH in the interstellar medium towards the quiescent 
molecular cloud G+0.693-0.027 located in the Galactic Center.
We have targeted the three groups of transitions from the $J$=2$-$1, 3$-$2, and 4$-$3 rotational lines, detecting 5 transitions that are unblended or only slightly blended.
The derived molecular abundance of NH$_2$OH is (2.1$\pm$0.9)$\times$10$^{-10}$. From the comparison of the derived abundance of NH$_2$OH and chemically related species, with those predicted by chemical models and measured in laboratory experiments, we favor the formation of NH$_2$OH in the interstellar medium via hydrogenation of NO on dust grain surfaces, with possibly a contribution of ice mantle NH$_3$ oxidation processes. Further laboratory studies and quantum chemical calculations are needed to completely rule out the formation of NH$_2$OH in the gas phase.

\end{abstract}


\keywords{Pre-biotic astrochemistry, Astrobiology, Interstellar molecules, Galactic center }




\section{Introduction} 
\label{sec:intro}

\begin{table*}
\centering
\tabcolsep 4.5pt
\caption{List of observed transitions of NH$_2$OH. We provide the parameters from the CDMS catalog entry 33503: frequencies, quantum numbers, base 10 logarithm of the integrated intensity at 300 K (log$I$), and upper state degeneracy (g$_{\rm u}$). We also show the values of $S_{\rm ul}\times\mu^2$, the base 10 logarithm of the Einstein coefficients (log $A_{\rm ul}$), and the energy of the upper levels ($E_{\rm u}$). The last column gives the information about the species with transitions blending with some of the NH$_2$OH lines.}
\begin{tabular}{ c c c c c c c l}
\hline
 Frequency & Transition  & log $I$ & S$_{\rm ul}\times\mu^2$ & log  $A_{\rm ul}$  & $g_{\rm u}$ & E$_{\rm u}$ &   Blending$^a$ \\
 (GHz) & ($J_{\rm K_a,K_c}$)  &   (nm$^2$ MHz) & (D$^2$) & (s$^{-1}$) & & (K) &   \\
\hline
  100.683580   & 2$_{1,2}-$1$_{1,1}$ & -4.8754  &   0.52036 &  -5.9078 &  5  & 15.2   & {\scriptsize blended with  CH$_3$OCHO} \\   
  100.748230   & 2$_{0,2}-$1$_{0,1}$ & -4.7384   &   0.69380 &  -5.7821 &  5  &   7.2    & {\scriptsize unblended} \\      
  100.807620   & 2$_{1,1}-$1$_{1,0}$ &  -4.8743 & 0.52041 &  -5.9062 & 5 &   15.2    & {\scriptsize blended with CH$_3$COOH and s-C$_2$H$_5$CHO} \\ 
  151.020701   & 3$_{1,3}-$2$_{1,2}$ &  -4.2821  & 0.92509 &  -5.2759 &  7 &  22.4     & {\scriptsize unblended} \\        
  151.101986   & 3$_{2,2}-$2$_{2,1}$ &  -4.5203  &  0.57820 &  -5.4793 &  7 &  46.3    & {\scriptsize blended with s-C$_2$H$_5$CHO  } \\       
  151.102324   & 3$_{2,1}-$2$_{2,0}$ &    -4.5203  &  0.57820  &  -5.4793 &  7 &  46.3     & {\scriptsize {\scriptsize blended with s-C$_2$H$_5$CHO  }} \\ 
  151.117668   & 3$_{0,3}-$2$_{0,2}$ &   -4.2189   &  1.04068  &  -5.2239 &  7  & 14.5    & {\scriptsize slightly blended with s-C$_2$H$_5$CHO$^{b}$} \\            
  151.207007   & 3$_{1,2}-$2$_{1,1}$ &  -4.2810    & 0.92520   &  -5.2742 &  7  & 22.5     & {\scriptsize slightly blended with unidentified line }\\       
  201.352578   & 4$_{1,4}-$3$_{1,3}$ &  -3.8964   &  1.30100  &  -4.8621 &  9 &  32.1    & {\scriptsize blended with CH$_3$SH and HDCO} \\   
  201.435072   & 4$_{3,2}-$3$_{3,1}$ &  -4.3192   & 0.60710   &  -5.1926 & 9 & 95.8    & {\scriptsize blended with CH$_3$OH}\\           
  201.435072   & 4$_{3,1}-$3$_{3,0}$ &  -4.3192   &   0.60710  &  -5.1926 &  9 & 95.8   & {\scriptsize blended with CH$_3$OH} \\       
  201.460663   & 4$_{2,3}-$3$_{2,2}$ &  -4.0274   & 1.04082 &  -4.9584 &  9 &  56.0    & {\scriptsize blended with CH$_3$OH}  \\        
  201.461598   & 4$_{2,2}-$3$_{2,1}$ &  -4.0274   &  1.04081 &  -4.9583 &  9 &  24.1    & {\scriptsize blended with CH$_3$OH} \\       
  201.481718   & 4$_{0,4}-$3$_{0,3}$ &  -3.8563   &  1.38777   &  -4.8333 &  9 &  32.1    & {\scriptsize slightly blended with unidentified line} \\            
  201.600715   & 4$_{1,3}-$3$_{1,2}$ &  -3.8954  &   1.30089 &  -4.8606 &  9 &  15.6    & {\scriptsize blended with H$_2$C$^{18}$O and CH$_3$CONH$_2$} \\  
\hline 
\end{tabular}
\label{tab:transitions}
{\\ (a) Detailed information about the species producing some blending is presented in Appendix \ref{appendix-b} (Table \ref{tab:table-contamination}). }
{\\ (b) The analysis of s-propanal (s-C$_2$H$_5$CHO) is presented in Appendix \ref{appendix-c}. }
\end{table*}

Life likely appeared about 3.8 billion years ago, 700 Myr after the formation of the Earth (\citealt{mojzsis1996}). However, the processes that allowed the transition from chemistry to biology are still uncertain.
Present-day life is mainly based on three biopolymers: RNA and DNA nucleic acids, which store genetic information, and proteins, responsible for most metabolic activities.
One of the main challenges for understanding the origin of life is the "chicken and egg" paradox: nucleic acids need catalytic proteins for their replication, while the biosynthesis of proteins requires the prior presence of genetic information coded in nucleic acids. 
To solve this problem, \citet{gilbert1986} proposed the hypothesis of a primordial RNA world, in which RNA molecules could have performed the functions that were carried out by DNA and proteins in the subsequent DNA/RNA/protein world.

One of the weaknesses of the RNA world hypothesis was
that ribonucleotides (RNA monomers) seemed difficult to be synthesized
under plausible prebiotic conditions (\citealt{atkins2011,ruiz-mirazo2014}). However, over the last years, this possibility 
has gained support thanks to recent prebiotic chemistry experiments that have shown that ribonucleotides can be formed from relatively simple organic molecules such as cyanoacetylene (HC$_3$N), cyanamide (NH$_2$CN), glycolaldehyde (CH$_2$OHCHO), urea (NH$_2$CONH$_2$) and/or hydroxylamine (NH$_2$OH; see  \citealt{powner2009,patel2015,becker2016,becker2019}).

Interestingly, except NH$_2$OH, all these RNA molecular precursors have been detected in the interstellar medium (ISM; \citealt{turner1971}, \citealt{turner1975}, \citealt{hollis_interstellar_2000}, \citealt{belloche2019}). As recently shown by \citet{becker2019}, NH$_2$OH is a key precursor in the unified synthesis of both pyrimidine and purine ribonucleotides.
However, observational searches of NH$_2$OH in the ISM have so far been unsuccessful (\citealt{pulliam2012,mcguire2015,ligterink2018}). This is a puzzling result since NH$_2$OH is expected to form efficiently on dust grains according to chemical models and laboratory experiments (\citealt{garrod_complex_2008,garrod2013ApJ,zheng2010,fedoseev2012,he2015}).

In this Letter, we present the first detection of NH$_2$OH in the ISM towards the quiescent molecular cloud G+0.693-0.027 (G+0.693, hereafter) located in the  Sgr B2 complex in the Galactic Center. This cloud shows a very rich chemistry in complex organic molecules  (\citealt{requena-torres_largest_2008,martin_tracing_2008,zeng2018,Rivilla2018,rivilla2019b}) including urea (\citealt{jimenez-serra2020}), another key ingredient in the pathway to ribonucleotides proposed by \citet{becker2019}. All this makes G+0.693 the perfect target where to search for NH$_2$OH in the ISM.

\begin{figure*}
\includegraphics[scale=0.227]{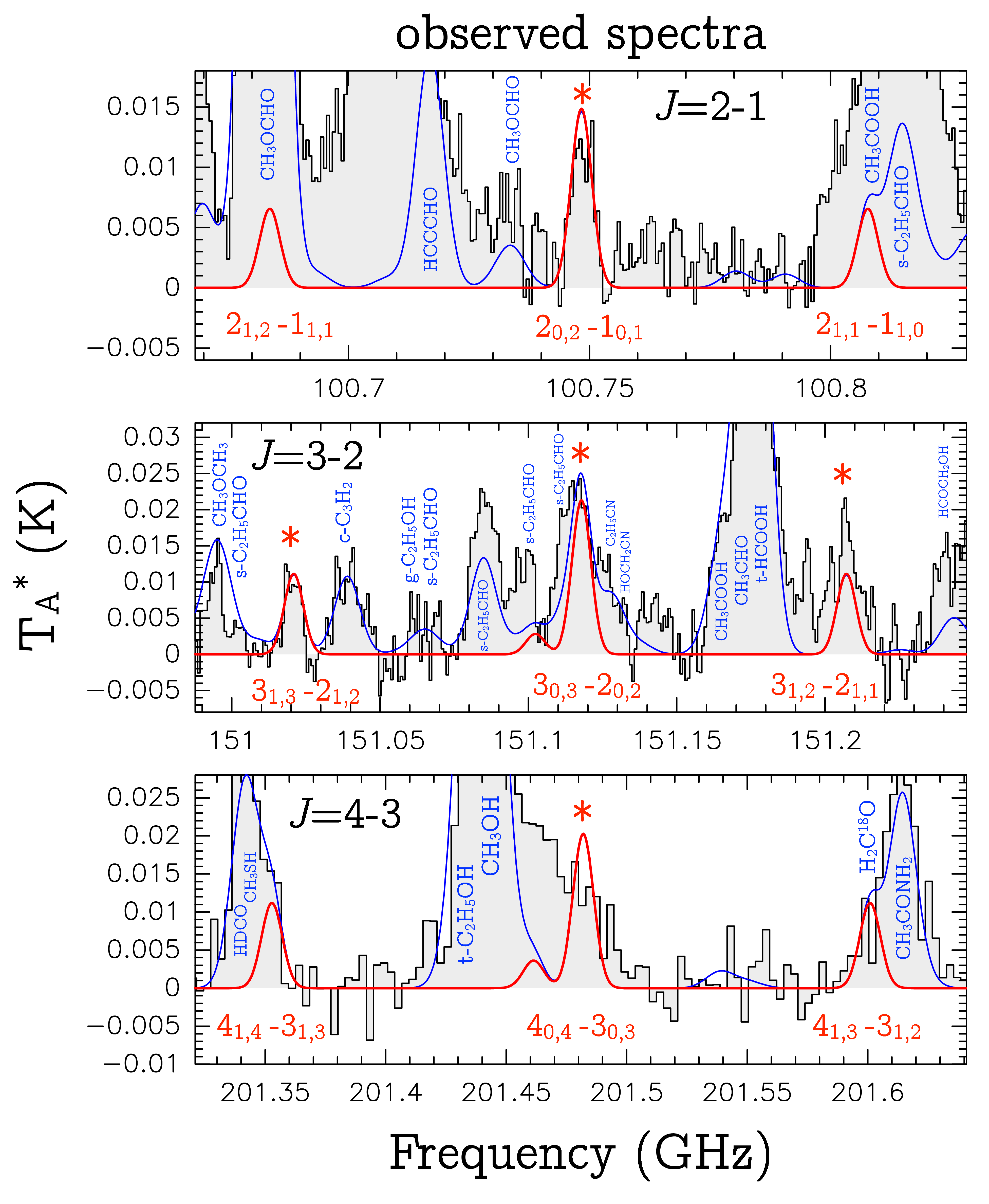}
\hspace{-0.2cm}
\includegraphics[scale=0.227]{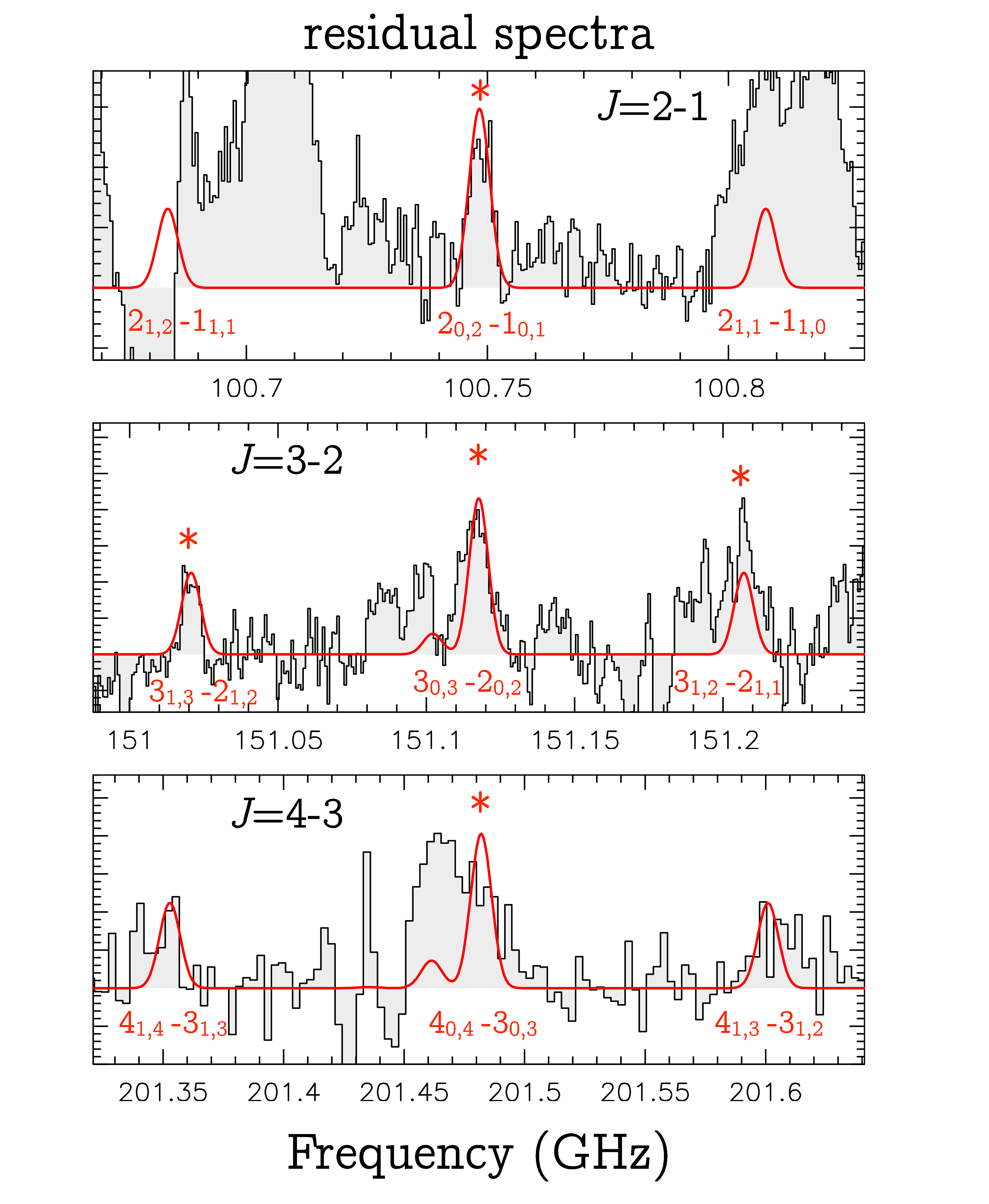}
\centering
\vspace{-3mm}
\caption{{\it Left panels:} Transitions of NH$_2$OH detected in the observed spectra of G+0.693. 
The nine brightest transitions are labelled with their quantum numbers. The five transitions unblended or only slightly blended are indicated with red asterisks.
The best LTE fit to the NH$_2$OH lines is shown in red, while the total contribution, including other molecular species (labelled) identified in the region is shown in blue.
{\it Right panels}: Residual spectra resulting from subtracting the LTE modelled contribution of all other identified molecular species but NH$_2$OH from the observed spectra. The best LTE fit to the NH$_2$OH lines is shown in red. 
}

\label{fig-spectra}
\end{figure*}

\section{Observations} 
\label{sec:observations}

We have carried out a high-sensitivity spectral survey at 3, 2 and 1 mm toward G+0.693 molecular cloud using the IRAM 30m telescope. The IRAM 30m observations were performed in three observing runs during 2019: April 10-16, August 13-19, and December 11-15 (projects numbers 172-18, 018-19 and 133-19). We used the broad-band Eight MIxer Receiver (EMIR) and the fast Fourier transform spectrometers in FTS200 mode, which provided a channel width of $\sim$200 kHz, i.e. a velocity resolution of $\sim$0.3$-$0.85 km s$^{-1}$. We smoothed the spectra to resolutions of 1.6, 1.4 and 4.0 km s$^{-1}$ for the 3, 2, and 1 mm data, respectively, high enough to sample the observed linewidths of $\sim$15$-$25 km s$^{-1}$, and to provide similar root-mean-square ($rms$) noise (2.6$-$3.8 mK), and optimal line visualization. We stress that the results of our analysis (described in Section \ref{sec:analysis}) are independent of the spectral resolution. Each frequency setup was observed several times shifting the central frequency by 20-100 MHz in order to identify possible spectral features resulting from unsuppressed image side band emission. The image-band lines, identified by comparing the same frequency range observed during two different spectral setups, were eliminated during the reduction of the data. The covered spectral ranges were 71.76$-$116.72$\,$GHz, 124.77$-$175.5$\,$GHz, 199.8$-$238.29 GHz, 252.52$-$260.30 GHz and 268.2$-$275.98 GHz. The observations were centered at $\alpha$(J2000.0)=$\,$17$^h$47$^m$22$^s$, $\delta$(J2000.0)=$\,-$28$^{\circ}$21$^{\prime}$27 $^{\prime\prime}$. The position switching mode was used in all observations with the off position located at (-885$^{\prime\prime}$,290$^{\prime\prime}$) from the source position. The intensity of the spectra is given in T$_A^*$ as the molecular emission toward G+0.693 is relatively extended over the beam (\citealt{requena-torres_organic_2006}; \citealt{Rivilla2018}).

\section{Analysis and Results} 
\label{sec:analysis}

The identification of the molecular lines was performed using the version 19/06/2020 of SLIM (Spectral Line Identification and Modeling) tool within the MADCUBA package{\footnote{Madrid Data Cube Analysis on ImageJ is a software developed at the Center of Astrobiology (CAB) in Madrid; http://cab.inta-csic.es/madcuba/Portada.html.}} (\citealt{martin2019}), which uses the spectroscopic entries from the Cologne Database for Molecular Spectroscopy (CDMS, \citealt{endres2016}) and the Jet Propulsion Laboratory (JPL; \citealt{Pickett1998}), and generates the synthetic spectra under the assumption of Local Thermodynamic Equilibrium (LTE) conditions.
The spectroscopic information of NH$_2$OH (CDMS entry 33503, from March 2003) was obtained from \citet{tsunekawa1972} and \citet{morino2000}. 
NH$_2$OH is nearly a prolate symmetric top molecule, in which the principal a- and c-axis are in a plane of symmetry.
Both a-type ($\Delta K_{\rm a}$=0) and c-type ($\Delta K_{\rm c}$=0) transitions are allowed, but only the former are expected to be detected in the ISM since the projected dipole moment along the a-axis is about ten times larger (\citealt{tsunekawa1972}).
The partition function of the ground vibrational state used in the analysis, obtained from the CDMS entry, does not have substantial contribution of the lowest-lying vibrational states (with energies $>$555 K, \citealt{morino2000}) due to the low excitation temperatures of the molecular emission towards G+0.693 \citep[T$_{ex}$$\sim$5$-$20$\,$K;][]{requena-torres_largest_2008,zeng2018,rivilla2019b}.

\begin{table*}
\centering
\tabcolsep 2.0pt
\caption{Derived physical parameters of NH$_2$OH and other molecular species that might be chemically related.}
\begin{tabular}{ c  c c c c c c  }
\hline
 Molecule & $N$   &  $T_{\rm ex}$ & v$_{\rm LSR}$ & $FWHM$  & Abundance$^a$ & Reference$^b$  \\
 & ($\times$10$^{13}$ cm$^{-2}$) & (K) & (km s$^{-1}$) & (km s$^{-1}$) & ($\times$10$^{-10}$) &   \\
\hline
NH$_2$OH  &   2.8$\pm$0.6 & 15  & 68.6$\pm$0.8 & 15 & 2.1$\pm$0.9 &  1  \\  
HNO & 5.8$\pm$1.3 & 3.7$\pm$0.2 & 66.4$\pm$0.4 & 20.5$\pm$0.9 & 4.3$\pm$1.8 &  1 \\ 
N$_2$O & 48$\pm$11 & 18$\pm$2 & 64.7$\pm$0.8  & 21$\pm$2 & 36$\pm$15 &  1 \\
NO & 3600$\pm$200 & 4.1$\pm$0.4  &  68$\pm$1  & 21$\pm$1 & 2700$\pm$700 & 2  \\
NO$_2$ & $<$29 & 10 & 69 & 20 & $<$22 & 1  \\ 
HONO &$<$0.2  & 10 & 69 & 20 & $<$0.16  &  1 \\ 
HNO$_3$  & $<$0.5 & 10 & 69 & 20 & $<$0.33 &  1 \\ 
\hline 
\end{tabular}
\label{tab:parameters}
{\\ (a) We adopted $N_{\rm H_2}$=(1.4$\pm$0.3)$\times$10$^{23}$ cm$^{-2}$, from \citet{martin_tracing_2008}.
(b) References: (1) This work; (2) \citet{zeng2018};
}
\end{table*}

We list in Table \ref{tab:transitions} all the a-type transitions of NH$_2$OH that fall in the observed data: 15 lines of the rotational transitions $J$=2$-$1, 3$-$2, and 4$-$3.
We note that transitions from higher energy levels ($J_{\rm u}>$5) have low predicted intensities due to the low excitation temperatures of G+0.693.
The fitted line profiles of NH$_2$OH (see details of the fit below) are shown with a red solid line in the left panels of Fig. \ref{fig-spectra}.
Five of the targeted NH$_2$OH transitions (1 at 3mm, 3 at 2mm, and 1 at 1mm, indicated with an asterisk in Fig. \ref{fig-spectra}), appear either unblended or only slightly blended.
The other transitions, which are blended with brighter lines from other molecules, have predicted intensities consistent with the observed spectra.

To properly evaluate the line contamination by other molecules we have searched for more than 300 different molecules in our dataset, including all the species detected and previously towards G+0.693 (e.g. \citealt{requena-torres_organic_2006,requena-torres_largest_2008,zeng2018,Rivilla2018, rivilla2019b,jimenez-serra2020}), and those detected so far in the ISM (\citealt{mcguire2018})\footnote{See also https://cdms.astro.uni-koeln.de/classic/molecules, and http://astrochymist.org/astrochymist$\_$mole.html}. 
We have fitted the emission of all the detected molecules, and produced the total LTE spectra, which is shown with a solid blue line in the left panels of Fig.$\,$\ref{fig-spectra}.
We list in the last column of Table \ref{tab:transitions}  the molecules with transitions that produce blending with some of the NH$_2$OH lines. More detailed information about all the identified molecular transitions that contribute to the line emission in the spectral ranges shown in Fig. \ref{fig-spectra} is presented in Appendix \ref{appendix-b}.

To highlight the emission from NH$_2$OH, we have computed the residual spectra obtained by subtracting from the observed spectra the predicted LTE contribution for all the identified molecular species but NH$_2$OH. The resulting residual spectra, shown in the right panels of Fig. \ref{fig-spectra}, clearly exhibit the spectral line profiles that match the LTE predictions for NH$_2$OH in red solid lines (see below), supporting the identification of  NH$_2$OH\footnote{We note that the NH$_2$OH transition at 100.683580 GHz is not well reproduced in the residual spectra due to the oversubtraction of the LTE model of two bright contaminating transitions of CH$_3$OCH$_3$: 9$_{0,9}-$8$_{0,8}$ A $\&$ E.}.

To derive the physical parameters that explain the NH$_2$OH emission, we used the SLIM-AUTOFIT tool of MADCUBA that provides the best non-linear least-squares LTE fit to the data using the Levenberg-Marquardt algorithm. The fitted parameters are: molecular column density ($N$), excitation temperature ($T_{\rm ex}$), central  velocity (v$_{\rm LSR}$) and full width half maximum ($FWHM$) of the Gaussian profiles. 
SLIM considers the solution of the radiative transfer as described in eq. 4 of \citet{martin2019}. For the modeling of G+0.693, where no continuum emission is detected (e.g. \citealt{ginsburg_dense_2016}), we consider $T_{\rm c}$=0, and the cosmic background temperature of $T_{\rm bg}$=2.73 K.
The fit to the NH$_2$OH profiles was performed by considering also the total contribution of the LTE emission from all the other identified molecules (blue line in Fig 1).
We fixed the linewidth to 15 km s$^{-1}$, which reproduces the profiles of the two strongest completely unblended transitions, 2$_{0,2}-$1$_{0,1}$ and 3$_{1,3}-$2$_{1,2}$ (Fig. \ref{fig-spectra}).
This value is consistent with the typical molecular linewidths found towards G+0.693, which span between 14 and 34 km s$^{-1}$ (e.g. \citealt{requena-torres_largest_2008,zeng2018}).
We considered typical values of $T_{\rm ex}$ found in G+0.693 in the range 5$-$20 K (see e.g. \citealt{zeng2018}). Low values (5$-$10 K) underestimate the observed intensities of the NH$_2$OH transitions at 2 and 1 mm, while values in the range 15$-$20 K are able to reproduce well all the transitions. We thus fixed $T_{\rm ex}$=15 K{\footnote The value of the partition function of NH$_2$OH, log $Q$(15K)=1.4613, is obtained by log-log linear interpolation of the values at 9.375 and 18.75 K from CDMS.} to perform the fit, noting that if the assumed fixed value is 20 K the derived column density would only vary by a factor of $\sim$1.1.
We then performed AUTOFIT, leaving column density and v$_{\rm LSR}$ as free parameters. The resulting best LTE fit is shown by the red curve in Fig. \ref{fig-spectra}, and the derived physical parameters are presented in Table \ref{tab:parameters}. To derive the final uncertainty of the column density we added quadratically the error derived from the fitting algorithm, and a 20$\%$ from the typical flux density uncertainty of the IRAM 30m telescope.
We obtain a column density of (2.8$\pm$0.6)$\times$10$^{13}$ cm$^{-2}$, which translates into a molecular abundance with respect to molecular hydrogen of (2.1$\pm$0.9)$\times$10$^{-10}$.

To gain insight into the interstellar chemistry and synthesis routes of NH$_2$OH in the ISM, we have also analysed the emission from chemically-related species. We detected the nitroxyl radical (HNO) and nitrous oxide (N$_2$O). The observed spectra along with the derived best LTE fits for these two molecules, reported for the first time in G+0.693 in this Letter, are presented in Appendix$\,$\ref{appendix}. Other species such as HONO, NO$_2$ and HNO$_3$ were not detected. 
The derived physical parameters for all these species are  reported in Table \ref{tab:parameters}. 

\begin{table*}
\centering
\caption{Molecular abundances and molecular abundances ratios.}
\begin{tabular}{c c c c c c c}
\hline
Region  & NH$_2$OH & NO   &  N$_2$O & NH$_2$OH/NO & NO/N$_2$O  & References$^a$ \\
\hline
G+0.693 & 2.1$\times$10$^{-10}$ & 2.7$\times$10$^{-7}$     & 3.5$\times$10$^{-9}$ & $\sim$8$\times$10$^{-4}$ & 76 & 1 \\
L1157-B1 & $<$1.4$\times$10$^{-8}$& (4-7)$\times$10$^{-6}$    & -  & $<$3$\times$10$^{-3}$ & - & 2,3 \\
IRAS 16293-2422 B & $<$3.1$\times$10$^{-11}$ & 1.7$\times$10$^{-9}$     & 3.3$\times$10$^{-9}$  & $<$2$\times$10$^{-2}$ & 0.5 & 4 \\
Sgr B2(N)$^b$  & $<$8$\times$10$^{-12}$ &  6.2$\times$10$^{-9}$ & 1.5$\times$10$^{-9}$ & $<$1.3$\times$10$^{-3}$  & 4 & 5,6 \\
Sgr B2(M)  & - &  9$\times$10$^{-9}$ & 1$\times$10$^{-9}$ & - & 9 & 7 \\
\hline
\end{tabular}
{\\ (a) References: (1) This work; (2) \citet{mcguire2015}; (3) \citet{codella2018}; (4) \citet{ligterink2018}; (5) \citet{pulliam2012} ; (6) \citet{Halfen2001}; (7) \citet{ziurys1994}.
(b) For a comparison with the NH$_2$OH upper limit obtained by \citet{pulliam2012}, the NO abundance has been recalculated using the column density reported by \citet{Halfen2001} and the hydrogen column density used by \citet{pulliam2012}.} 
\label{tab:comparison}
\end{table*}

\section{Discussion}
\label{sec:discussion}

\subsection{Formation of NH$_2$OH in the ISM}

Table \ref{tab:comparison} compares the abundance of NH$_2$OH in G+0.693 with the upper limits found in other regions where it has been previously searched for. We find that the abundance of NH$_2$OH in G+0.693 is about a factor of 6 and 25 higher than the upper limits measured in the hot corino IRAS 16293$-$2422 B and the hot core Sgr B2(N), respectively (\citealt{ligterink2018,pulliam2012}). Unfortunately, the upper limit toward the protostellar shock L1157-B1 (\citealt{mcguire2015}) is too high for a meaningful comparison. 

The several routes that have been proposed for the synthesis of NH$_2$OH in the ISM, all of them based on surface chemistry, are
summarized in Table$\,$\ref{tab:routes}. 
The hydrogenation of nitrogen oxide (NO) (Route 1; Table$\,$\ref{tab:routes}) was suggested theoretically by \citet{charnley2001} and it has been confirmed experimentally by different groups \citep{fedoseev2012,congiu2012a}. In this process, the successive hydrogenation of NO and its products HNO and H$_2$NO, is barrierless (or it occurs with very small activation barriers), yielding NH$_2$OH in a very efficient way even at low temperatures \citep{congiu2012a,fedoseev2012}. This chemical scheme requires the presence of significant amounts of NO on grain surfaces. For the case of G+0.693, this is not an issue since high abundances of this molecule are measured toward this source (of $\sim$3$\times$10$^{-7}$; Table$\,$\ref{tab:comparison}). 
Similar high abundances of NO have been measured toward the shocked region B1 in the L1157 molecular outflow (\citealt{codella2018}, Table \ref{tab:comparison}). These authors showed that such high NO abundance can be produced via the neutral-neutral reaction N + OH $\rightarrow$ NO + H in the gas phase. Thus, a significant fraction of NO would then be available to deplete onto dust grains at the low dust temperatures of G+0.693 \citep[$T\leq$30 K;][]{Rodriguez-Fernandez2004}.

Although NO could be destroyed prior to be depleted onto dust grains due to its conversion to N$_2$O via the gas-phase reaction NO + NH $\rightarrow$  N$_2$O+H \citep{Halfen2001}, this mechanism does not seem efficient in G+0.693, where the NO/N$_2$O ratio of 76 is much higher than those found in hotter sources (Table \ref{tab:comparison}). 
As indicated by \citet{Halfen2001}, this gas phase reaction is only efficient for kinetic temperatures $>$150 K. 
The gas kinetic temperature in G+0.693 is 50$-$120 K (\citealt{huettemeister_kinetic_1993,Krieger2017}), i.e. lower than in hot cores/corinos (with temperatures of 200$-$300 K; \citealt{belloche_complex_2013,Jorgensen2016}), which would explain the low efficiency of the gas-phase conversion of NO into N$_2$O. Note, however, that N$_2$O  
could also be formed on the surface of dust grains (\citealt{fedoseev2012,congiu2012a}), and be subsequently desorbed by shocks in G+0.693, which might explain the observed  N$_2$O. 

In addition, we have also detected HNO (see Table \ref{tab:parameters} and Appendix \ref{appendix}), the first hydrogenation product of NO, supporting this chemical route. Although several gas-phase formation routes of HNO exist, such as e.g. NH$_2$ + O $\rightarrow$ H + HNO, or HCO + NO  $\rightarrow$ HNO + CO (see the KIDA database\footnote{http://kida.astrophy.u-bordeaux.fr}, \citealt{wakelam2012}), \citet{hasegawa1992} showed that
the predicted abundance of HNO in models that consider only gas-phase reactions is of just a few 10$^{-11}$.
Since the derived abundance of HNO in G+0.693 is about one order of magnitude higher, the formation of HNO in G+0.693 likely occurs mainly on the surface of dust grains through the NO hydrogenation chain that also produces NH$_2$OH.
 
Other NH$_2$OH formation pathways have been proposed by laboratory experiments (\citealt{nishi1984,zheng2010}) and chemical modeling (\citealt{garrod2013ApJ}) involving the amidogen radical NH$_2$ (see Routes 2a and 2b in Table$\,$\ref{tab:routes}). 
However, we note again that the low dust temperature in G+0.693 ($T\leq$30 K, \citealt{Rodriguez-Fernandez2004}) would limit the mobility of heavier radicals such as NH$_2$ or OH on grain surfaces (Routes 2a and 2b) compared to atomic H (Route 1), which suggests that the hydrogenation route is more plausible.

The third grain-surface reaction for the formation of NH$_2$OH is the oxidation of NH$_3$ (Routes 3 and 4 in Table \ref{tab:routes}). The experiments of \citet{he2015} and \citet{tsegaw2017} show that NH$_2$OH can be formed on the surface of dust grains from NH$_3$ and O/O$_2$ at warm ($\sim$70 K) and cold temperatures ($<$10 K), respectively. In the \citet{he2015} work, however, the predicted ice-mantle abundance of NH$_2$OH (with respect to H) lies in the range 10$^{-5}-$10$^{-6}$, which is several orders of magnitude higher than the one observed in the gas phase of G+0.693. 
Besides missing destruction pathways, this could be due to an optimistic value of the reaction rate of NH$_3$ + O $\rightarrow$ NH$_2$OH in \citet{he2015}, since the oxidation of NH$_3$ in their experiments occurs via the Eley-Rideal mechanism, which may be more efficient than the Langmuir-Hinschelwood mechanism taking place on interstellar dust grains.

The second oxidation route, NH$_3$ + O$_2$ $\rightarrow$ NH$_2$OH + O, has been studied by the laboratory experiments by \citet{tsegaw2017}, in which interstellar ice analogs were irradiated with energetic electrons to mimic the effect of cosmic rays. Since G+0.693 is affected by a high cosmic-ray ionization rate (\citealt{requena-torres_organic_2006,zeng2018}), this oxidation route might be a possible additional mechanism for the formation of NH$_2$OH.

In contrast to grain-surface chemistry, very little is known about possible gas-phase formation routes for NH$_2$OH. Experimental works have shown that gas-phase production of NH$_2$OH through the reaction between HNO$^+$ and H$_2$, is inefficient (\citealt{blagojevic2003}).
Neutral-neutral reactions such as NH$_3$ + OH or NH$_2$ + OH are expected to be efficient at the kinetic temperatures of G+0.693 (see KIDA database). However, these reactions mainly yield, respectively, NH$_2$ + H$_2$O and NH + H$_2$O. Indeed, no NH$_2$OH seems to be found in the products of the gas-phase reaction NH$_3$ + H$_2$O (Prof. D. Heard, private communication), although further experiments and quantum chemical calculations are needed to confirm this result.

\begin{table}
\centering
\tabcolsep 3.0pt
\caption{Grain-surface routes proposed for the formation of NH$_2$OH.}
\begin{tabular}{l c c}
\hline
& Route & References \\
\hline
1) & NO $\rightarrow$ HNO $\rightarrow$ H$_2$NO/HNOH $\rightarrow$ NH$_2$OH & i \\
2a) & NH$_2$ + H + H$_2$O $\rightarrow$ NH$_2$OH +H$_2$ & ii \\
2b) & NH$_2$ + OH $\rightarrow$ NH$_2$OH & iii \\
3) & NH$_3$ +O $\rightarrow$  NH$_2$OH  & iv \\
4) & NH$_3$ +O$_2$ $\rightarrow$  NH$_2$OH + O  & v \\
\hline
\end{tabular}
\label{tab:routes}
{\\ (i) \citet{fedoseev2012,congiu2012a}; 
(ii) \citet{nishi1984};
(iii) \citet{nishi1984,zheng2010};
(iv) \citet{he2015};
(v) \citet{tsegaw2017}}  .
\end{table}

\subsection{The role of NH$_2$OH in prebiotic chemistry}

Recent prebiotic experiments conducted under plausible early Earth conditions have shown that NH$_2$OH is a central species in the synthesis of pyrimidine rubonucleotides (\citealt{becker2019}), through its reaction with HC$_3$N in the presence of urea (NH$_2$CONH$_2$). HC$_3$N is very abundant in G+0.693 (\citealt{zeng2018}), and NH$_2$CONH$_2$ has also been detected recently toward this source (\citealt{jimenez-serra2020}). Thus, the only missing ingredient was NH$_2$OH, reported in this paper. Precursors of purine ribonucleotides such as cyanamide (NH$_2$CN) and glycolaldehyde (CH$_2$OHCHO)  (\citealt{powner2009,becker2016}), have also been detected in G+0.693 (\citealt{requena-torres_largest_2008,zeng2018}). Therefore, the G+0.693 molecular cloud presents all precursors of ribonucleotides. 

In summary, the detection of NH$_2$OH in the molecular cloud G+0.693 supports the idea that essential prebiotic precursors of the RNA world could have been synthesized initially in the ISM, to be subsequently transferred to small Solar-System bodies, and later on to young planets through the impact of meteorites and comets (see e.g. \citealt{chyba1992,pierazzo1999,pearce2015,martins2018,pizzarello2006,rubin2019,rivilla2020}).

\bibliography{bibliography}{}

\begin{thebibliography}{}
\expandafter\ifx\csname natexlab\endcsname\relax\def\natexlab#1{#1}\fi
\providecommand{\url}[1]{\href{#1}{#1}}
\providecommand{\dodoi}[1]{doi:~\href{http://doi.org/#1}{\nolinkurl{#1}}}
\providecommand{\doeprint}[1]{\href{http://ascl.net/#1}{\nolinkurl{http://ascl.net/#1}}}
\providecommand{\doarXiv}[1]{\href{https://arxiv.org/abs/#1}{\nolinkurl{https://arxiv.org/abs/#1}}}

\bibitem[{{Atkins} {et~al.}(2011){Atkins}, {Gesteland}, \& {Cech}}]{atkins2011}
{Atkins}, J., {Gesteland}, R., \& {Cech}, T.~. 2011, {66 pages, 73 col \& 23
  b/w illus}, ed. C.~S. H.~L. Press, 366

\bibitem[{Becker {et~al.}(2016)Becker, Thoma, Deutsch, Gehrke, Mayer, Zipse, \&
  Carell}]{becker2016}
Becker, S., Thoma, I., Deutsch, A., {et~al.} 2016, Science, 352, 833,
  \dodoi{10.1126/science.aad2808}

\bibitem[{Becker {et~al.}(2019)Becker, Feldmann, Wiedemann, Okamura, Schneider,
  Iwan, Crisp, Rossa, Amatov, \& Carell}]{becker2019}
Becker, S., Feldmann, J., Wiedemann, S., {et~al.} 2019, Science, 366, 76,
  \dodoi{10.1126/science.aax2747}

\bibitem[{{Belloche} {et~al.}(2019){Belloche}, {Garrod}, {M{\"u}ller},
  {Menten}, {Medvedev}, {Thomas}, \& {Kisiel}}]{belloche2019}
{Belloche}, A., {Garrod}, R.~T., {M{\"u}ller}, H.~S.~P., {et~al.} 2019, \aap,
  628, A10, \dodoi{10.1051/0004-6361/201935428}

\bibitem[{Belloche {et~al.}(2013)Belloche, M\"uller, Menten, Schilke, \&
  Comito}]{belloche_complex_2013}
Belloche, A., M\"uller, H. S.~P., Menten, K.~M., Schilke, P., \& Comito, C.
  2013, Astronomy and Astrophysics, 559, A47,
  \dodoi{10.1051/0004-6361/201321096}

\bibitem[{{Blagojevic} {et~al.}(2003){Blagojevic}, {Petrie}, \&
  {Bohme}}]{blagojevic2003}
{Blagojevic}, V., {Petrie}, S., \& {Bohme}, D.~K. 2003, \mnras, 339, L7,
  \dodoi{10.1046/j.1365-8711.2003.06351.x}

\bibitem[{{Charnley} {et~al.}(2001){Charnley}, {Rodgers}, \&
  {Ehrenfreund}}]{charnley2001}
{Charnley}, S.~B., {Rodgers}, S.~D., \& {Ehrenfreund}, P. 2001, \aap, 378,
  1024, \dodoi{10.1051/0004-6361:20011193}

\bibitem[{Chyba \& Sagan(1992)}]{chyba1992}
Chyba, C., \& Sagan, C. 1992, Nature, 355, 125, \dodoi{10.1038/355125a0}

\bibitem[{{Codella} {et~al.}(2018){Codella}, {Viti}, {Lefloch}, {Holdship},
  {Bachiller}, {Bianchi}, {Ceccarelli}, {Favre}, {Jim{\'e}nez-Serra}, {Podio},
  \& {Tafalla}}]{codella2018}
{Codella}, C., {Viti}, S., {Lefloch}, B., {et~al.} 2018, \mnras, 474, 5694,
  \dodoi{10.1093/mnras/stx3196}

\bibitem[{{Congiu} {et~al.}(2012){Congiu}, {Fedoseev}, {Ioppolo}, {Dulieu},
  {Chaabouni}, {Baouche}, {Lemaire}, {Laffon}, {Parent}, {Lamberts}, {Cuppen},
  \& {Linnartz}}]{congiu2012a}
{Congiu}, E., {Fedoseev}, G., {Ioppolo}, S., {et~al.} 2012, \apjl, 750, L12,
  \dodoi{10.1088/2041-8205/750/1/L12}

\bibitem[{{Endres} {et~al.}(2016){Endres}, {Schlemmer}, {Schilke}, {Stutzki},
  \& {M{\"u}ller}}]{endres2016}
{Endres}, C.~P., {Schlemmer}, S., {Schilke}, P., {Stutzki}, J., \&
  {M{\"u}ller}, H. S.~P. 2016, Journal of Molecular Spectroscopy, 327, 95,
  \dodoi{10.1016/j.jms.2016.03.005}

\bibitem[{{Fedoseev} {et~al.}(2012){Fedoseev}, {Ioppolo}, {Lamberts}, {Zhen},
  {Cuppen}, \& {Linnartz}}]{fedoseev2012}
{Fedoseev}, G., {Ioppolo}, S., {Lamberts}, T., {et~al.} 2012, \jcp, 137,
  054714, \dodoi{10.1063/1.4738893}

\bibitem[{{Garrod}(2013)}]{garrod2013ApJ}
{Garrod}, R.~T. 2013, \apj, 765, 60, \dodoi{10.1088/0004-637X/765/1/60}

\bibitem[{Garrod {et~al.}(2008)Garrod, Widicus~Weaver, \&
  Herbst}]{garrod_complex_2008}
Garrod, R.~T., Widicus~Weaver, S.~L., \& Herbst, E. 2008, The Astrophysical
  Journal, 682, 283, \dodoi{10.1086/588035}

\bibitem[{Gilbert(1986)}]{gilbert1986}
Gilbert, W. 1986, Nature, 319, 618, \dodoi{10.1038/319618a0}

\bibitem[{Ginsburg {et~al.}(2016)Ginsburg, Henkel, Ao, Riquelme, Kauffmann,
  Pillai, Mills, Requena-Torres, Immer, Testi, Ott, Bally, Battersby, Darling,
  Aalto, Stanke, Kendrew, Kruijssen, Longmore, Dale, Guesten, \&
  Menten}]{ginsburg_dense_2016}
Ginsburg, A., Henkel, C., Ao, Y., {et~al.} 2016, Astronomy \& Astrophysics,
  586, A50, \dodoi{10.1051/0004-6361/201526100}

\bibitem[{{Halfen} {et~al.}(2001){Halfen}, {Apponi}, \& {Ziurys}}]{Halfen2001}
{Halfen}, D.~T., {Apponi}, A.~J., \& {Ziurys}, L.~M. 2001, \apj, 561, 244,
  \dodoi{10.1086/322770}

\bibitem[{{Hasegawa} {et~al.}(1992){Hasegawa}, {Herbst}, \&
  {Leung}}]{hasegawa1992}
{Hasegawa}, T.~I., {Herbst}, E., \& {Leung}, C.~M. 1992, \apjs, 82, 167,
  \dodoi{10.1086/191713}

\bibitem[{{He} {et~al.}(2015){He}, {Vidali}, {Lemaire}, \& {Garrod}}]{he2015}
{He}, J., {Vidali}, G., {Lemaire}, J.-L., \& {Garrod}, R.~T. 2015, \apj, 799,
  49, \dodoi{10.1088/0004-637X/799/1/49}

\bibitem[{Hollis {et~al.}(2000)Hollis, Lovas, \&
  Jewell}]{hollis_interstellar_2000}
Hollis, J.~M., Lovas, F.~J., \& Jewell, P.~R. 2000, The Astrophysical Journal
  Letters, 540, L107, \dodoi{10.1086/312881}

\bibitem[{H\"uettemeister {et~al.}(1993)H\"uettemeister, Wilson, Bania, \&
  Mart\'in-Pintado}]{huettemeister_kinetic_1993}
H\"uettemeister, S., Wilson, T.~L., Bania, T.~M., \& Mart\'in-Pintado, J. 1993,
  Astronomy and Astrophysics, 280, 255.
\newblock \url{http://adsabs.harvard.edu/abs/1993A%26A...280..255H}

\bibitem[{Jim\'enez-Serra {et~al.}(2020)Jim\'enez-Serra, Mart\'n-Pintado,
  Rivilla, Rodr\'iguez-Almeida, Alonso~Alonso, Zeng, Cocinero, Mart\'in,
  Requena-Torres, Mart\'in-Domenech, \& Testi}]{jimenez-serra2020}
Jim\'enez-Serra, I., Mart\'n-Pintado, J., Rivilla, V.~M., {et~al.} 2020,
  Astrobiology, 0, null, \dodoi{10.1089/ast.2019.2125}

\bibitem[{{J{\o}rgensen} {et~al.}(2016){J{\o}rgensen}, {van der Wiel},
  {Coutens}, {Lykke}, {M{\"u}ller}, {van Dishoeck}, {Calcutt}, {Bjerkeli},
  {Bourke}, {Drozdovskaya}, {Favre}, {Fayolle}, {Garrod}, {Jacobsen},
  {{\"O}berg}, {Persson}, \& {Wampfler}}]{Jorgensen2016}
{J{\o}rgensen}, J.~K., {van der Wiel}, M.~H.~D., {Coutens}, A., {et~al.} 2016,
  \aap, 595, A117, \dodoi{10.1051/0004-6361/201628648}

\bibitem[{{Krieger} {et~al.}(2017){Krieger}, {Ott}, {Beuther}, {Walter},
  {Kruijssen}, {Meier}, {Mills}, {Contreras}, {Edwards}, {Ginsburg}, {Henkel},
  {Henshaw}, {Jackson}, {Kauffmann}, {Longmore}, {Mart{\'{\i}}n}, {Morris},
  {Pillai}, {Rickert}, {Rosolowsky}, {Shinnaga}, {Walsh}, {Yusef-Zadeh}, \&
  {Zhang}}]{Krieger2017}
{Krieger}, N., {Ott}, J., {Beuther}, H., {et~al.} 2017, The Astrophysical
  Journal, 850, 77, \dodoi{10.3847/1538-4357/aa951c}

\bibitem[{{Ligterink} {et~al.}(2018){Ligterink}, {Calcutt}, {Coutens},
  {Kristensen}, {Bourke}, {Drozdovskaya}, {M{\"u}ller}, {Wampfler}, {van der
  Wiel}, {van Dishoeck}, \& {J{\o}rgensen}}]{ligterink2018}
{Ligterink}, N.~F.~W., {Calcutt}, H., {Coutens}, A., {et~al.} 2018, \aap, 619,
  A28, \dodoi{10.1051/0004-6361/201731980}

\bibitem[{{Mart{\'\i}n} {et~al.}(2019){Mart{\'\i}n}, {Mart{\'\i}n-Pintado},
  {Blanco-S{\'a}nchez}, {Rivilla}, {Rodr{\'\i}guez-Franco}, \&
  {Rico-Villas}}]{martin2019}
{Mart{\'\i}n}, S., {Mart{\'\i}n-Pintado}, J., {Blanco-S{\'a}nchez}, C.,
  {et~al.} 2019, \aap, 631, A159, \dodoi{10.1051/0004-6361/201936144}

\bibitem[{Mart\'in {et~al.}(2008)Mart\'in, Requena-Torres, Mart\'in-Pintado, \&
  Mauersberger}]{martin_tracing_2008}
Mart\'in, S., Requena-Torres, M.~A., Mart\'in-Pintado, J., \& Mauersberger, R.
  2008, The Astrophysical Journal, 678, 245, \dodoi{10.1086/533409}

\bibitem[{Martins(2018)}]{martins2018}
Martins, Z. 2018, Life, 8, \dodoi{10.3390/life8030028}

\bibitem[{{McGuire}(2018)}]{mcguire2018}
{McGuire}, B.~A. 2018, \apjs, 239, 17, \dodoi{10.3847/1538-4365/aae5d2}

\bibitem[{{McGuire} {et~al.}(2015){McGuire}, {Carroll}, {Dollhopf}, {Crockett},
  {Corby}, {Loomis}, {Burkhardt}, {Shingledecker}, {Blake}, \&
  {Remijan}}]{mcguire2015}
{McGuire}, B.~A., {Carroll}, P.~B., {Dollhopf}, N.~M., {et~al.} 2015, \apj,
  812, 76, \dodoi{10.1088/0004-637X/812/1/76}

\bibitem[{{Mojzsis} {et~al.}(1996){Mojzsis}, {Arrhenius}, {McKeegan},
  {Harrison}, {Nutman}, \& {Friend}}]{mojzsis1996}
{Mojzsis}, S.~J., {Arrhenius}, G., {McKeegan}, K.~D., {et~al.} 1996, Nature,
  384, 55–59, \dodoi{10.1038/384055a0}

\bibitem[{Morino {et~al.}(2000)Morino, Yamada, Klein, Belov, Winnewisser,
  Bocquet, Wlodarczak, Lodyga, \& Kreglewski}]{morino2000}
Morino, I., Yamada, K., Klein, H., {et~al.} 2000, Journal of Molecular
  Structure, 517-518, 367 ,
  \dodoi{https://doi.org/10.1016/S0022-2860(99)00263-X}

\bibitem[{{Nishi} {et~al.}(1984){Nishi}, {Shinohara}, \& {Okuyama}}]{nishi1984}
{Nishi}, N., {Shinohara}, H., \& {Okuyama}, T. 1984, \jcp, 80, 3898,
  \dodoi{10.1063/1.447172}

\bibitem[{{Patel}(2015)}]{patel2015}
{Patel}, B.~H., P. C. R. D. J. D. C. D. S. J.~D. 2015, Nature Chemistry, 7,
  301, \dodoi{10.1038/nchem.2202}

\bibitem[{Pearce \& Pudritz(2015)}]{pearce2015}
Pearce, B. K.~D., \& Pudritz, R.~E. 2015, The Astrophysical Journal, 807, 85,
  \dodoi{10.1088/0004-637x/807/1/85}

\bibitem[{{Pickett} {et~al.}(1998){Pickett}, {Poynter}, {Cohen}, {Delitsky},
  {Pearson}, \& {M{\"u}ller}}]{Pickett1998}
{Pickett}, H.~M., {Poynter}, R.~L., {Cohen}, E.~A., {et~al.} 1998, \jqsrt, 60,
  883, \dodoi{10.1016/S0022-4073(98)00091-0}

\bibitem[{Pierazzo \& Chyba(1999)}]{pierazzo1999}
Pierazzo, E., \& Chyba, C.~F. 1999, Meteoritics \& Planetary Science, 34, 909,
  \dodoi{10.1111/j.1945-5100.1999.tb01409.x}

\bibitem[{{Pizzarello} {et~al.}(2006){Pizzarello}, {Cooper}, \&
  {Flynn}}]{pizzarello2006}
{Pizzarello}, S., {Cooper}, G.~W., \& {Flynn}, G.~J. 2006, {The Nature and
  Distribution of the Organic Material in Carbonaceous Chondrites and
  Interplanetary Dust Particles}, ed. D.~S. {Lauretta} \& H.~Y. {McSween}, 625

\bibitem[{Powner(2009)}]{powner2009}
Powner, Matthew~W., G. B. S. J.~D. 2009, Nature, 459, 239–242,
  \dodoi{10.1038/nature08013}

\bibitem[{{Pulliam} {et~al.}(2012){Pulliam}, {McGuire}, \&
  {Remijan}}]{pulliam2012}
{Pulliam}, R.~L., {McGuire}, B.~A., \& {Remijan}, A.~J. 2012, \apj, 751, 1,
  \dodoi{10.1088/0004-637X/751/1/1}

\bibitem[{Requena-Torres {et~al.}(2008)Requena-Torres, Mart\'in-Pintado,
  Mart\'in, \& Morris}]{requena-torres_largest_2008}
Requena-Torres, M.~A., Mart\'in-Pintado, J., Mart\'in, S., \& Morris, M.~R.
  2008, The Astrophysical Journal, 672, 352, \dodoi{10.1086/523627}

\bibitem[{Requena-Torres {et~al.}(2006)Requena-Torres, Mart\'in-Pintado,
  Rodr\'iguez-Franco, Mart\'in, Rodr\'iguez-Fern\'andez, \&
  de~Vicente}]{requena-torres_organic_2006}
Requena-Torres, M.~A., Mart\'in-Pintado, J., Rodr\'iguez-Franco, A., {et~al.}
  2006, Astronomy \& Astrophysics, 455, 971, \dodoi{10.1051/0004-6361:20065190}

\bibitem[{{Rivilla} {et~al.}(2018){Rivilla}, {Jim{\'e}nez-Serra}, {Zeng},
  {Mart{\'{\i}}n}, {Mart{\'{\i}}n-Pintado}, {Armijos-Abenda{\~n}o}, {Viti},
  {Aladro}, {Riquelme}, {Requena-Torres}, {Qu{\'e}nard}, {Fontani}, \&
  {Beltr{\'a}n}}]{Rivilla2018}
{Rivilla}, V.~M., {Jim{\'e}nez-Serra}, I., {Zeng}, S., {et~al.} 2018, Monthly
  Notices of the Royal Astronomical Society, \dodoi{10.1093/mnrasl/slx208}

\bibitem[{{Rivilla} {et~al.}(2019){Rivilla}, {Mart{\'\i}n-Pintado},
  {Jim{\'e}nez-Serra}, {Zeng}, {Mart{\'\i}n}, {Armijos-Abenda{\~n}o},
  {Requena-Torres}, {Aladro}, \& {Riquelme}}]{rivilla2019b}
{Rivilla}, V.~M., {Mart{\'\i}n-Pintado}, J., {Jim{\'e}nez-Serra}, I., {et~al.}
  2019, \mnras, 483, L114, \dodoi{10.1093/mnrasl/sly228}

\bibitem[{{Rivilla} {et~al.}(2020){Rivilla}, {Drozdovskaya}, {Altwegg},
  {Caselli}, {Beltr{\'a}n}, {Fontani}, {van der Tak}, {Cesaroni}, {Vasyunin},
  {Rubin}, {Lique}, {Marinakis}, {Testi}, {Rosina Team}, {Balsiger},
  {Berthelier}, {de Keyser}, {Fiethe}, {Fuselier}, {Gasc}, {Gombosi},
  {S{\'e}mon}, \& {Tzou}}]{rivilla2020}
{Rivilla}, V.~M., {Drozdovskaya}, M.~N., {Altwegg}, K., {et~al.} 2020, \mnras,
  492, 1180, \dodoi{10.1093/mnras/stz3336}

\bibitem[{{Rodr{\'{\i}}guez-Fern{\'a}ndez}
  {et~al.}(2004){Rodr{\'{\i}}guez-Fern{\'a}ndez}, {Mart{\'{\i}}n-Pintado},
  {Fuente}, \& {Wilson}}]{Rodriguez-Fernandez2004}
{Rodr{\'{\i}}guez-Fern{\'a}ndez}, N.~J., {Mart{\'{\i}}n-Pintado}, J., {Fuente},
  A., \& {Wilson}, T.~L. 2004, Astronomy and Astrophysics, 427, 217,
  \dodoi{10.1051/0004-6361:20041370}

\bibitem[{Rubin {et~al.}(2019)Rubin, Bekaert, Broadley, Drozdovskaya, \&
  Wampfler}]{rubin2019}
Rubin, M., Bekaert, D.~V., Broadley, M.~W., Drozdovskaya, M.~N., \& Wampfler,
  S.~F. 2019, ACS Earth and Space Chemistry, 3, 1792,
  \dodoi{10.1021/acsearthspacechem.9b00096}

\bibitem[{{Ruiz-Mirazo} {et~al.}(2014){Ruiz-Mirazo}, {Briones}, \& {de la
  Escosura}}]{ruiz-mirazo2014}
{Ruiz-Mirazo}, K., {Briones}, C., \& {de la Escosura}, A. 2014, Chemical
  reviews, 114, 285, \dodoi{10.1021/cr2004844}

\bibitem[{{Tsegaw} {et~al.}(2017){Tsegaw}, {G{\'o}bi}, {F{\"o}rstel},
  {Maksyutenko}, {Sander}, \& {Kaiser}}]{tsegaw2017}
{Tsegaw}, Y.~A., {G{\'o}bi}, S., {F{\"o}rstel}, M., {et~al.} 2017, Journal of
  Physical Chemistry A, 121, 7477, \dodoi{10.1021/acs.jpca.7b07500}

\bibitem[{Tsunekawa(1972)}]{tsunekawa1972}
Tsunekawa, S. 1972, Journal of the Physical Society of Japan, 33, 167,
  \dodoi{10.1143/JPSJ.33.167}

\bibitem[{{Turner}(1971)}]{turner1971}
{Turner}, B.~E. 1971, \apjl, 163, L35, \dodoi{10.1086/180662}

\bibitem[{{Turner} {et~al.}(1975){Turner}, {Liszt}, {Kaifu}, \&
  {Kisliakov}}]{turner1975}
{Turner}, B.~E., {Liszt}, H.~S., {Kaifu}, N., \& {Kisliakov}, A.~G. 1975,
  \apjl, 201, L149, \dodoi{10.1086/181963}

\bibitem[{{Wakelam} {et~al.}(2012){Wakelam}, {Herbst}, {Loison}, {Smith},
  {Chandrasekaran}, {Pavone}, {Adams}, {Bacchus-Montabonel}, {Bergeat},
  {B{\'e}roff}, {Bierbaum}, {Chabot}, {Dalgarno}, {van Dishoeck}, {Faure},
  {Geppert}, {Gerlich}, {Galli}, {H{\'e}brard}, {Hersant}, {Hickson},
  {Honvault}, {Klippenstein}, {Le Picard}, {Nyman}, {Pernot}, {Schlemmer},
  {Selsis}, {Sims}, {Talbi}, {Tennyson}, {Troe}, {Wester}, \&
  {Wiesenfeld}}]{wakelam2012}
{Wakelam}, V., {Herbst}, E., {Loison}, J.-C., {et~al.} 2012, The Astrophysical
  Journal Supplement, 199, 21, \dodoi{10.1088/0067-0049/199/1/21}

\bibitem[{{Zeng} {et~al.}(2018){Zeng}, {Jim{\'e}nez-Serra}, {Rivilla},
  {Mart{\'{\i}}n}, {Mart{\'{\i}}n-Pintado}, {Requena-Torres},
  {Armijos-Abenda{\~n}o}, {Riquelme}, \& {Aladro}}]{zeng2018}
{Zeng}, S., {Jim{\'e}nez-Serra}, I., {Rivilla}, V.~M., {et~al.} 2018, Monthly
  Notices of the Royal Astronomical Society, 478, 2962,
  \dodoi{10.1093/mnras/sty1174}

\bibitem[{{Zheng} \& {Kaiser}(2010)}]{zheng2010}
{Zheng}, W., \& {Kaiser}, R.~I. 2010, Journal of Physical Chemistry A, 114,
  5251, \dodoi{10.1021/jp911946m}

\bibitem[{{Ziurys} {et~al.}(1994){Ziurys}, {Apponi}, {Hollis}, \&
  {Snyder}}]{ziurys1994}
{Ziurys}, L.~M., {Apponi}, A.~J., {Hollis}, J.~M., \& {Snyder}, L.~E. 1994,
  \apjl, 436, L181, \dodoi{10.1086/187662}

\end{thebibliography}
\bibliographystyle{aasjournal}

\section{ACKNOWLEDGEMENTS} 

We acknowledge the anonymous reviewers for their careful reading of the manuscript and their useful comments.
This work is based on observations carried out under projects number 172-18, 018-19 and 133-19 with the IRAM 30m telescope. IRAM is supported by INSU/CNRS (France), MPG (Germany) and IGN (Spain). 
We thank the IRAM-30m staff for the precious help during the different observing runs.
V.M.R. has received funding from the European Union's Horizon 2020 research and innovation programme under the Marie Sk\l{}odowska-Curie grant agreement No 664931. I.J.-S. and J.M.-P. have received partial support from the Spanish FEDER (project number ESP2017-86582-C4-1-R), the Ministry of Science and Innovation through project number PID2019-105552RB-C41, and State Research Agency (AEI) through project number MDM-2017-0737 Unidad de Excelencia Mar\'ia de Maeztu$-$Centro de Astrobiolog\'ia (INTA-CSIC).

\clearpage

\begin{appendices}

\clearpage

\section{Full molecular identification}
\label{appendix-b}
 
We list in Table \ref{tab:table-contamination} the identified molecular transitions of the molecules that contribute to the spectra shown in Fig. \ref{fig-spectra}. We have searched for more than 300 different molecules in our spectral survey, including all the species detected previously towards G+0.693 (e.g. \citealt{requena-torres_organic_2006,requena-torres_largest_2008,zeng2018,Rivilla2018, rivilla2019b, jimenez-serra2020}), and those detected so far in the ISM\footnote{See also https://cdms.astro.uni-koeln.de/classic/molecules, and http://astrochymist.org/astrochymist$\_$mole.html} (\citealt{mcguire2018}). 
We show in Table \ref{tab:table-contamination} transitions with line intensities $I_{\rm L}\geq$ 1 mK.

\begin{table}[h]
\tabcolsep 5.0pt
\caption{Transitions of the molecules that contribute to the spectra shown in Fig. \ref{fig-spectra}. We present the name of the species, the frequency of the transition, and the line intensity $I_{\rm L}$.}
\begin{minipage}{0.31\textwidth}
\begin{tabular}{c c c}
\hline
 Molecule & Freq.   &  $I_{\rm L}$    \\
 & (GHz) &  (mK)     \\
\hline
\multicolumn{3}{c}{3 mm - NH$_2$OH $J$=2$-$1} \\
\hline
CH$_3$OCHO   & 100.6815450   &   93.3   \\ 
CH$_3$OCHO   & 100.6833680    & 93.4   \\  
NH$_2$OH  & 100.6835800   &  6.1    \\  
CH$_3$OCHO  & 100.6931270   &   1.22   \\  
HCCCHO  & 100.7169300  &  18.4  \\ 
CH$_3$OCHO   & 100.734805  & 1.1   \\  
NH$_2$OH   & 100.7482300  & 12.1   \\  
NH$_2$OH  &  100.8076200 & 6.1   \\   
CH$_3$COOH  & 100.8119422  & 1.52   \\
s-C$_2$H$_5$CHO  & 100.8149861  &  6.3  \\   
s-C$_2$H$_5$CHO    &  100.8149861 & 6.3   \\   
    &  &    \\ 
     &  &    \\ 
      &  &    \\ 
    &  &    \\ 
     &  &    \\ 
      &  &    \\ 
       &  &    \\ 
        &  &    \\ 
    &  &    \\ 
    &  &    \\ 
    &  &    \\ 
    &  &    \\ 
    &  &    \\   
    &  &    \\ 
    &  &    \\ 
    &  &    \\ 
    &  &    \\ 
    &  &    \\ 
    &  &    \\   
    &  &    \\ 
    &  &    \\ 
    &  &    \\ 
    &  &    \\ 
    &  &    \\ 
    &  &    \\   
    &  &    \\ 
    &  &    \\ 
    &  &    \\ 
    &  &    \\ 
    &  &    \\ 
    &  &    \\   
    &  &    \\ 
    &  &    \\ 
\end{tabular}
\end{minipage} 
\begin{minipage}{0.34\textwidth}
\begin{tabular}{c c c}
\hline
Molecule & Freq.   &  $I_{\rm L}$    \\
 & (GHz) &  (mK)     \\
\hline
\multicolumn{3}{c}{2 mm - NH$_2$OH $J$=3$-$2} \\
\hline
CH$_3$OCH$_3$ & 150.992109      &   1.8   \\
CH$_3$OCH$_3$ & 150.9921790      &   2.8   \\
s-C$_2$H$_5$CHO & 150.9952334      &  1.5    \\
CH$_3$OCH$_3$ & 150.9953910      &   7.4   \\ 
s-C$_2$H$_5$CHO & 150.9954363      &  1.5    \\
CH$_3$OCH$_3$ &150.9954363     &  1.5   \\  
CH$_3$OCH$_3$  &  150.9986390  &  4.6  \\
NH$_2$OH  & 151.0207010   &   11.6   \\  
c-C$_3$H$_2$  & 151.0391730   &  3.2  \\
g-C$_2$H$_5$OH  & 151.0641006   &  2.2   \\
s-C$_2$H$_5$CHO  &  151.0653970  &  1.1   \\  
s-C$_2$H$_5$CHO      & 151.0654207   &  1.1    \\
s-C$_2$H$_5$CHO  & 151.0831633   &   1.1   \\
s-C$_2$H$_5$CHO  & 151.0831861   &  1.1    \\
s-C$_2$H$_5$CHO    &  151.0970518  &  1.1 \\
s-C$_2$H$_5$CHO    &  151.0970740  & 1.1 \\
NH$_2$OH    & 151.1019860   &  2.2 \\
NH$_2$OH    &  151.1023240  & 2.2 \\
s-C$_2$H$_5$CHO  & 151.1148181   & 1.1 \\
s-C$_2$H$_5$CHO  & 151.1148394   & 1.1 \\
NH$_2$OH    &  151.1176680 & 19.4 \\
C$_2$H$_5$CN  & 151.1272640   & 3.5 \\
HOCH$_2$CN   &  151.1300319  & 1.0 \\
CH$_3$COOH & 151.1620321   & 1.5  \\
CH$_3$CHO   &  151.1671882  & 18.4 \\
t-HCOOH   & 151.1762475   & 81.4 \\
NH$_2$OH &  151.2070070  & 11.6 \\
HCOCH$_2$OH & 151.2430164  & 4.5 \\
  &  &    \\ 
 &   &  \\
 &   &  \\
 &   &  \\
 &   &  \\
  &  &    \\ 
 &   &  \\
 &   &  \\
 &   &  \\
 &   &  \\
 &   &  \\
 &   &  \\
 &   &  \\
 &   &  \\
 &   &  \\
 &   &  \\
\end{tabular}
\end{minipage} 
\begin{minipage}{0.33\textwidth}
\begin{tabular}{c c c}
\hline
Molecule & Freq.   &  $I_{\rm L}$    \\
 & (GHz) &  (K)     \\
\hline
\multicolumn{3}{c}{1 mm - NH$_2$OH $J$=4$-$3} \\
\hline
HDCO  &  201.3413620  &  13.2    \\
CH$_3$SH    & 201.3440150    &   14.6  \\  
NH$_2$OH    & 201.3525783   &   13.6   \\ 
t-C$_2$H$_5$OH      &  201.4372959  &  24.3  \\  
CH$_3$OH  &  201.4454903  &   77.0   \\
NH$_2$OH  & 201.4606626   &  3.3   \\
NH$_2$OH      & 201.4615977   &   3.3   \\  
NH$_2$OH   & 201.4817175   & 21.6\\
NH$_2$OH   & 201.6007146   &  13.6    \\
H$_2$C$^{18}$O   & 201.6142649   &  21.4    \\
CH$_3$CONH$_2$  & 201.6210000   &   1.7   \\
    &  &    \\ 
        &  &    \\ 
    &  &    \\ 
    &  &    \\ 
    &  &    \\
    &  &    \\ 
    &  &    \\ 
    &  &    \\   
    &  &    \\
    &  &    \\ 
    &  &    \\ 
    &  &    \\ 
    &  &    \\ 
    &  &    \\ 
    &  &    \\ 
    &  &    \\ 
    &  &    \\ 
    &  &    \\   
    &  &    \\ 
    &  &    \\
     &  &    \\ 
    &  &    \\ 
    &  &    \\   
    &  &    \\ 
    &  &    \\ 
    &  &    \\ 
    &  &    \\ 
    &  &    \\ 
    &  &    \\   
    &  &    \\ 
    &  &    \\
    &  &    \\ 
    &  &    \\   
\end{tabular}
\end{minipage} \hfill
\label{tab:table-contamination}
\end{table}

\clearpage

\section{Detection of Propanal (s-C$_2$H$_5$CHO) towards G+0.693}
\label{appendix-c}

\begin{figure*}
\includegraphics[scale=0.225,angle=0]{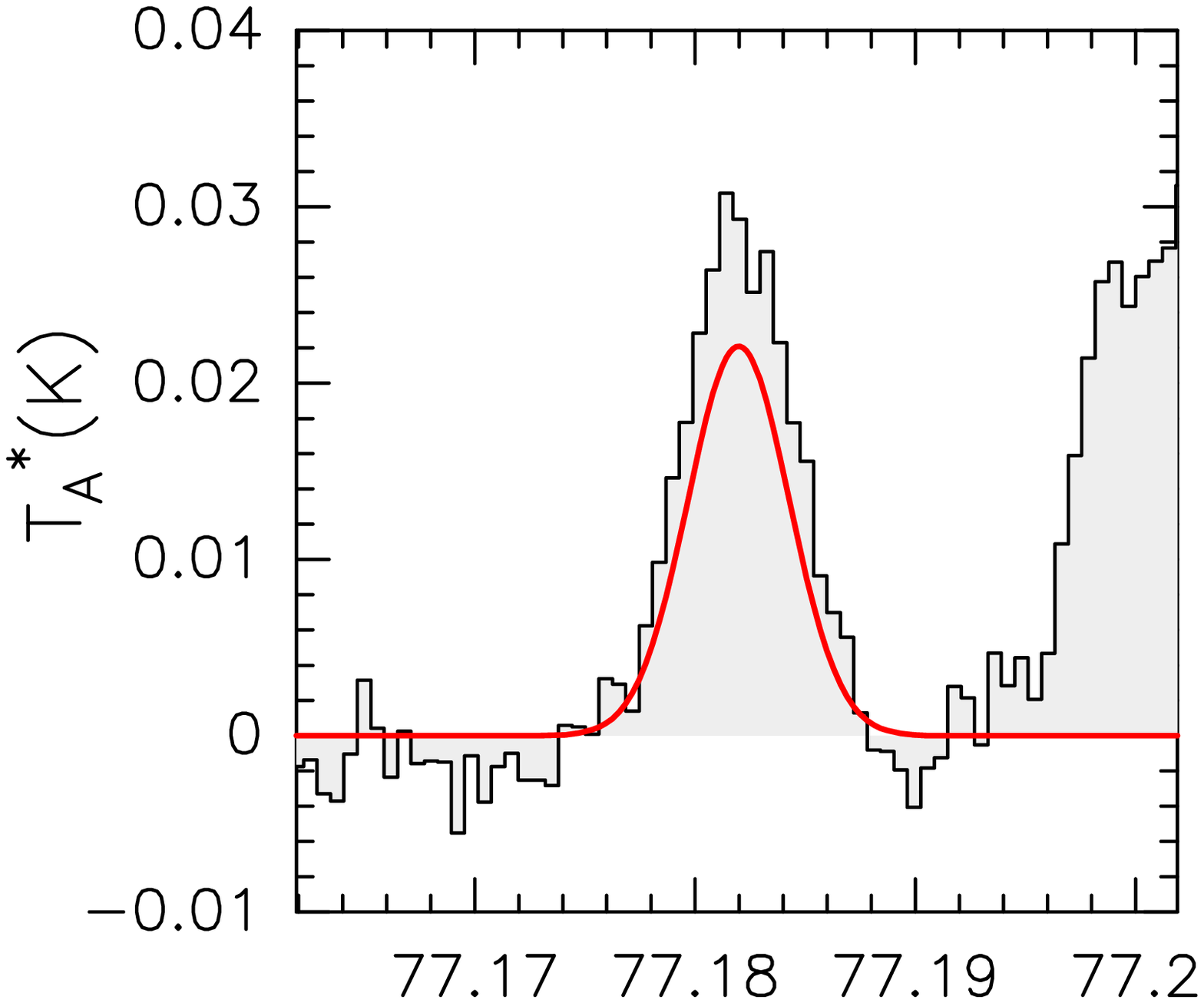}
\hspace{1mm}
\includegraphics[scale=0.225,angle=0]{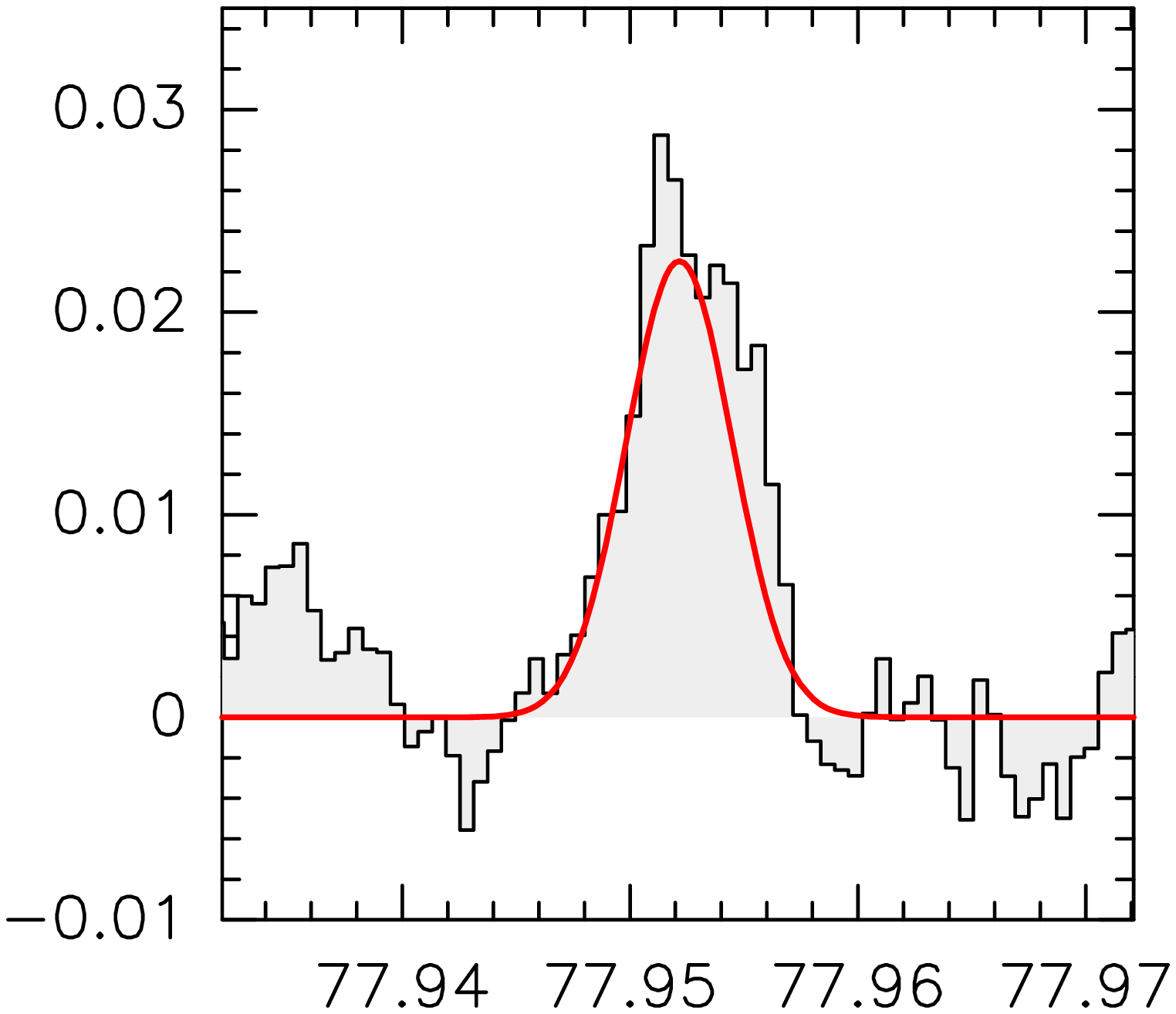}
\hspace{1mm}
\includegraphics[scale=0.225,angle=0]{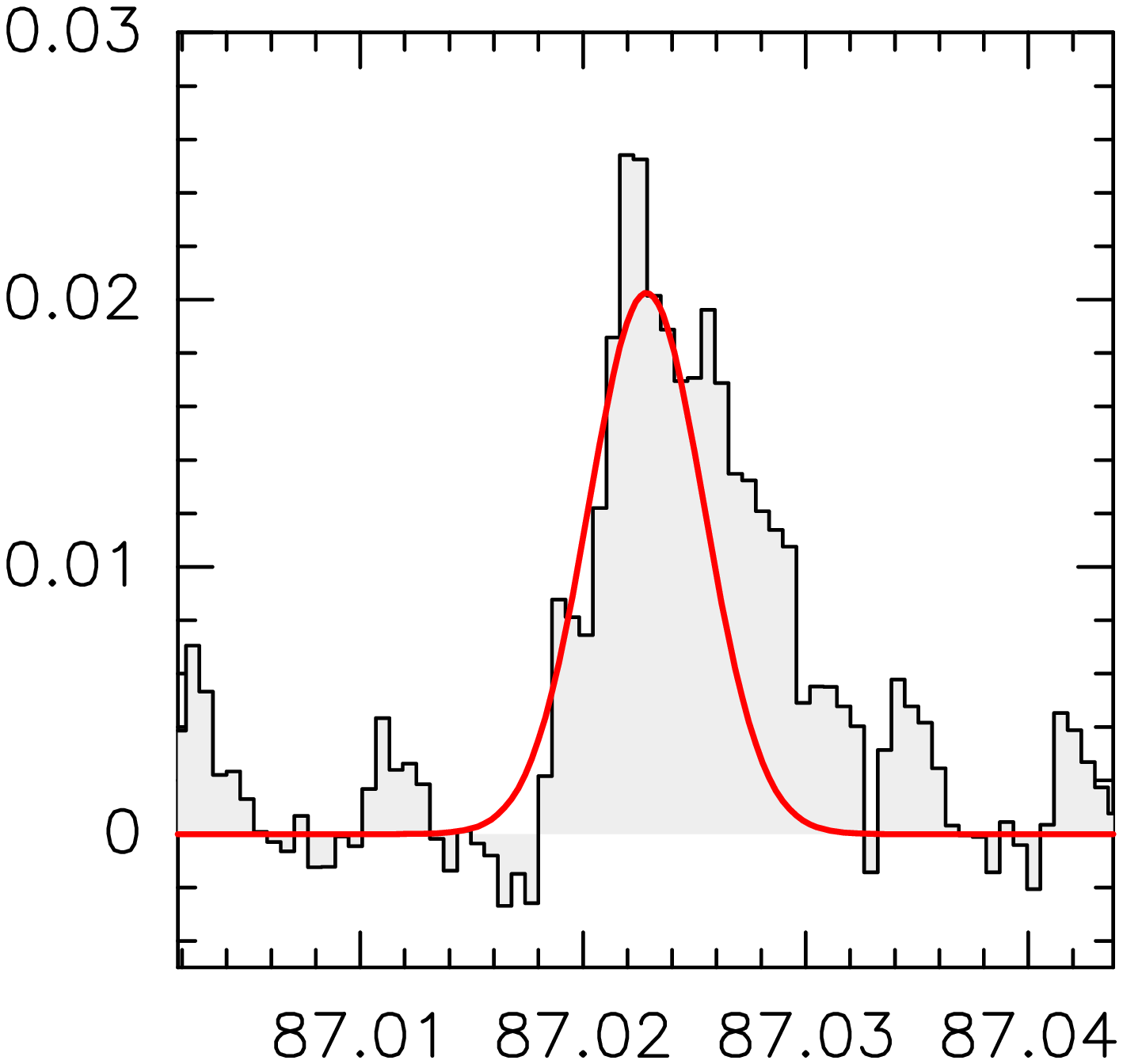}
\hspace{1mm}
\includegraphics[scale=0.225,angle=0]{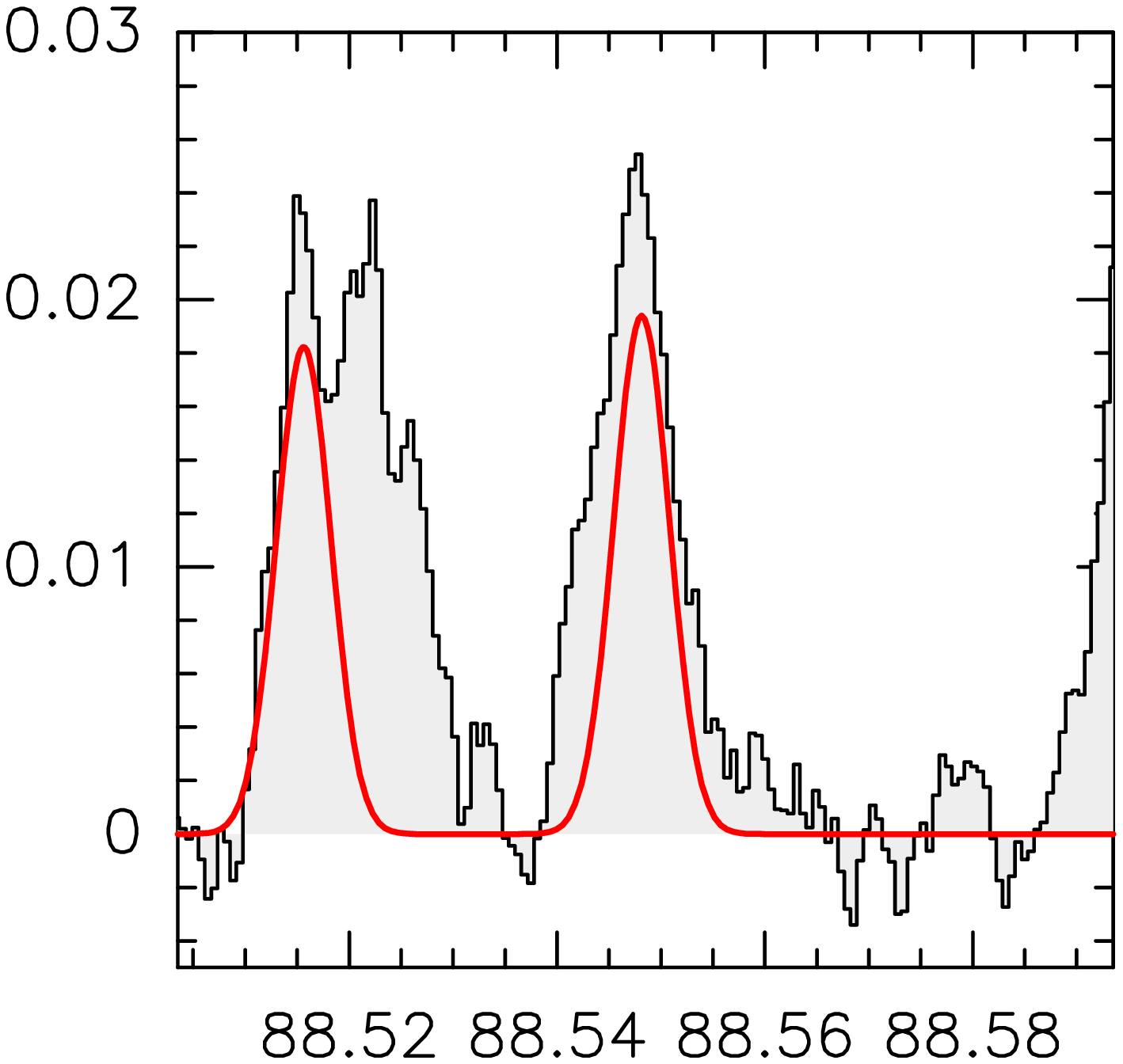}
\hspace{1mm}
\includegraphics[scale=0.225,angle=0]{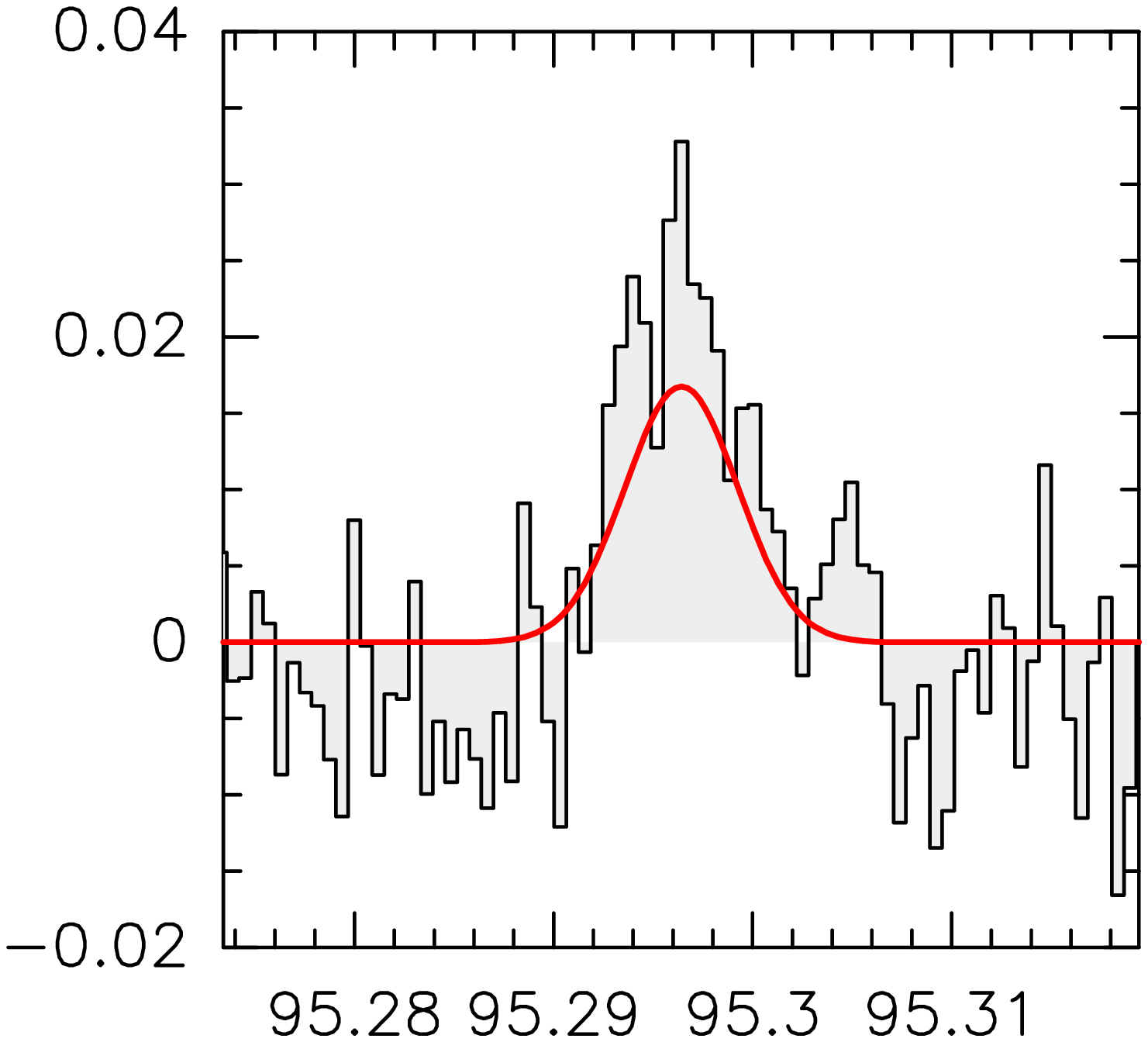}
\vskip5mm
\includegraphics[scale=0.225,angle=0]{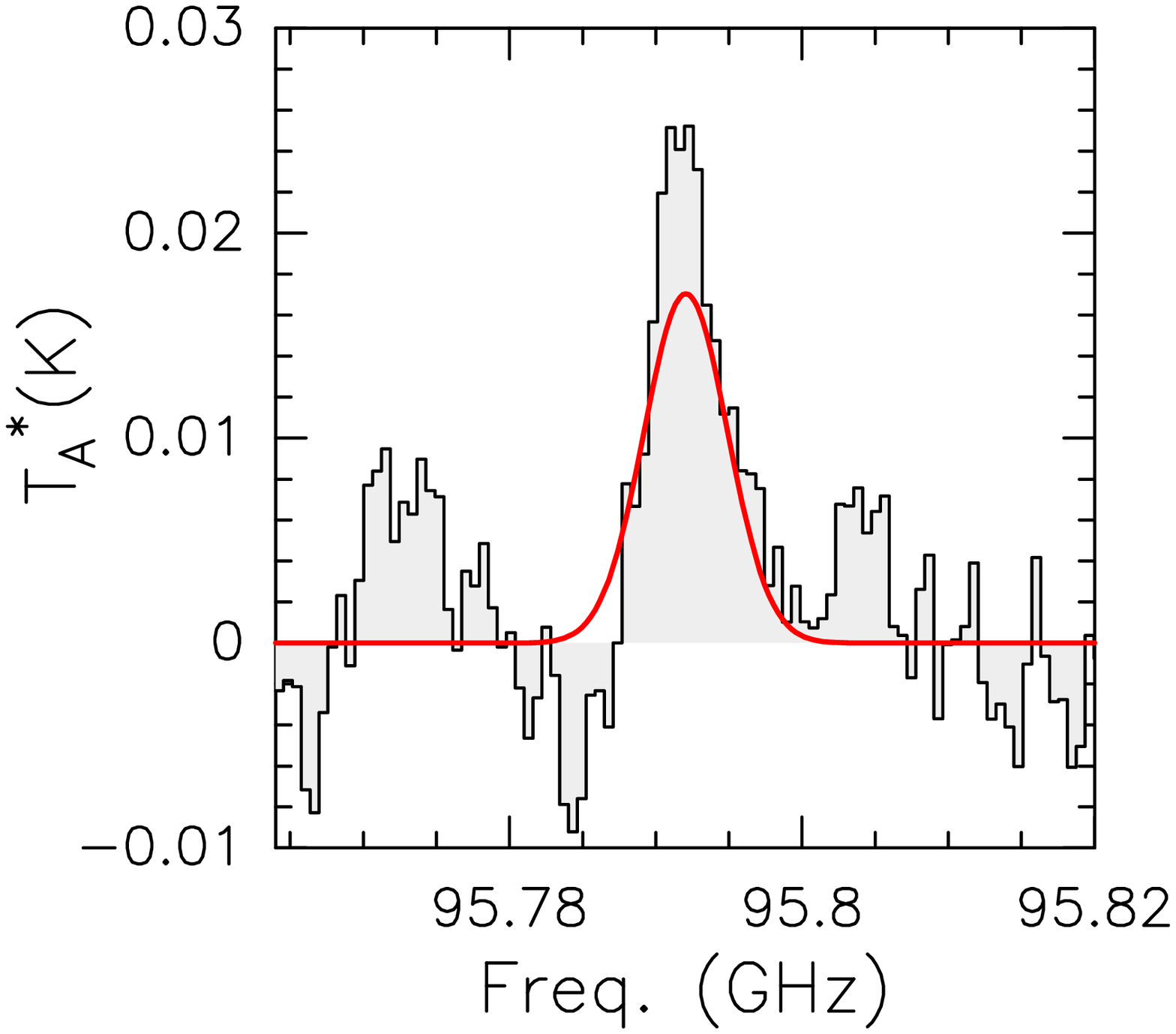}
\hspace{2mm}
\includegraphics[scale=0.225,angle=0]{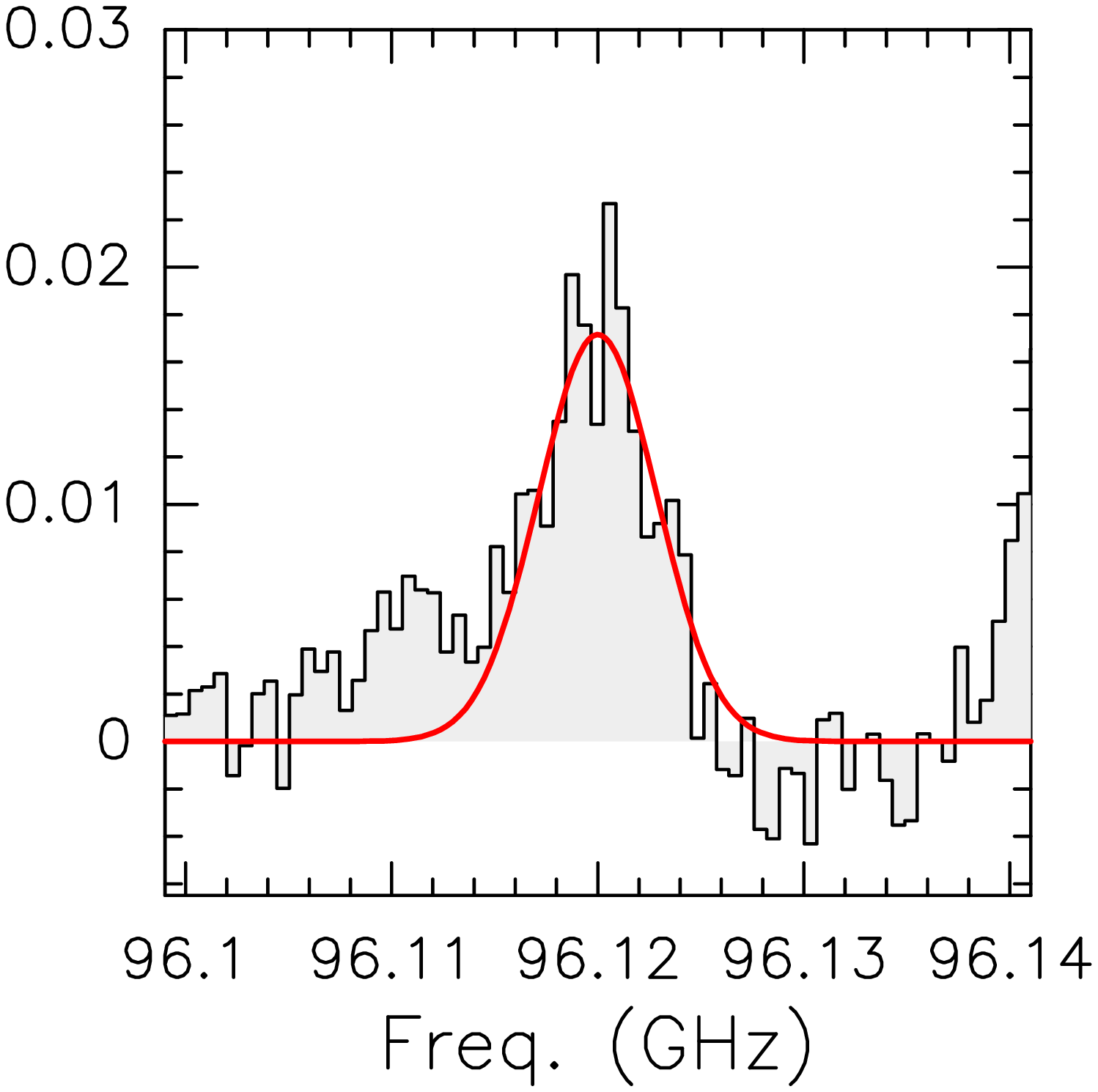}
\hspace{2mm}
\includegraphics[scale=0.225,angle=0]{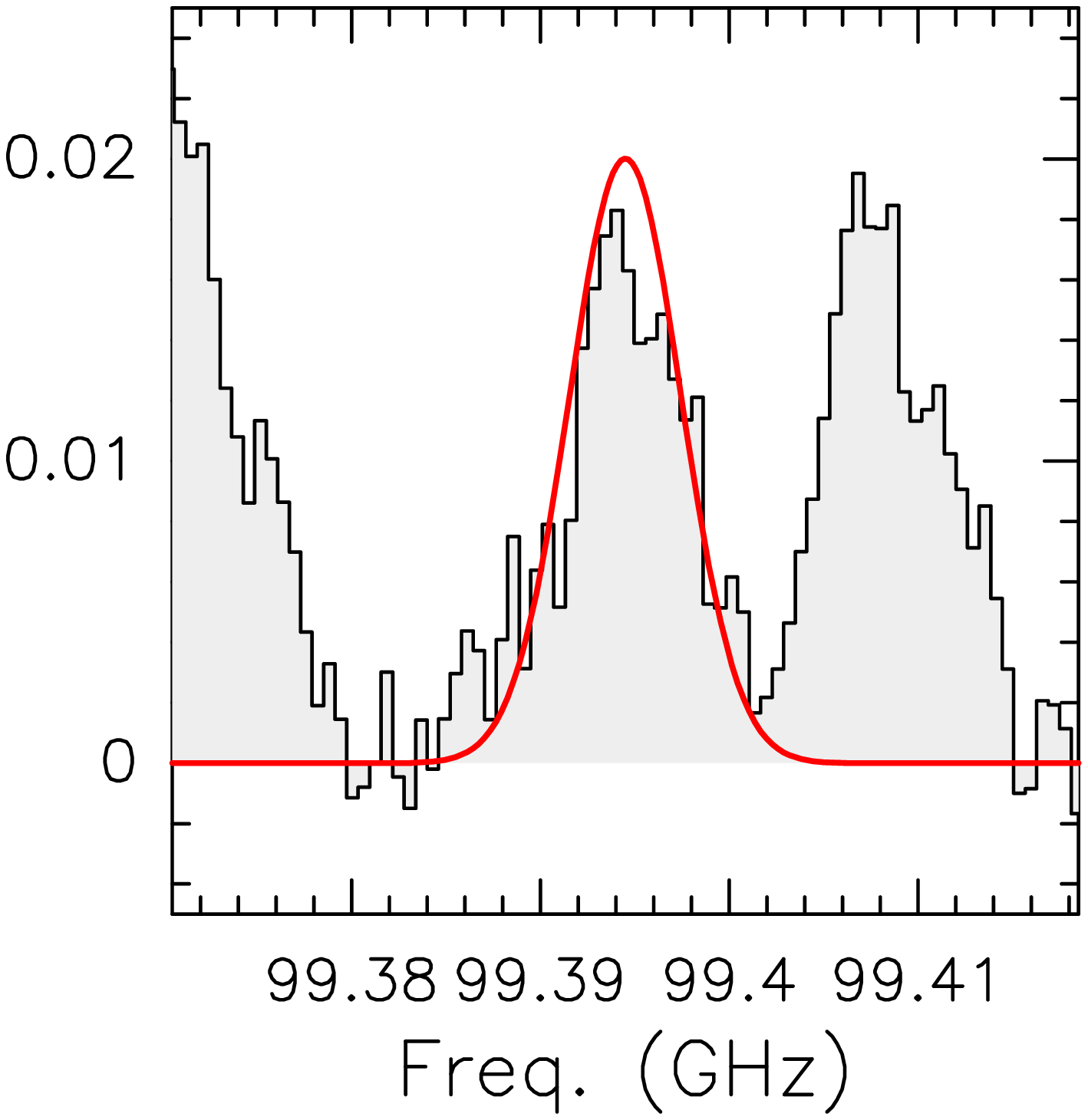}
\hspace{2mm}
\includegraphics[scale=0.225,angle=0]{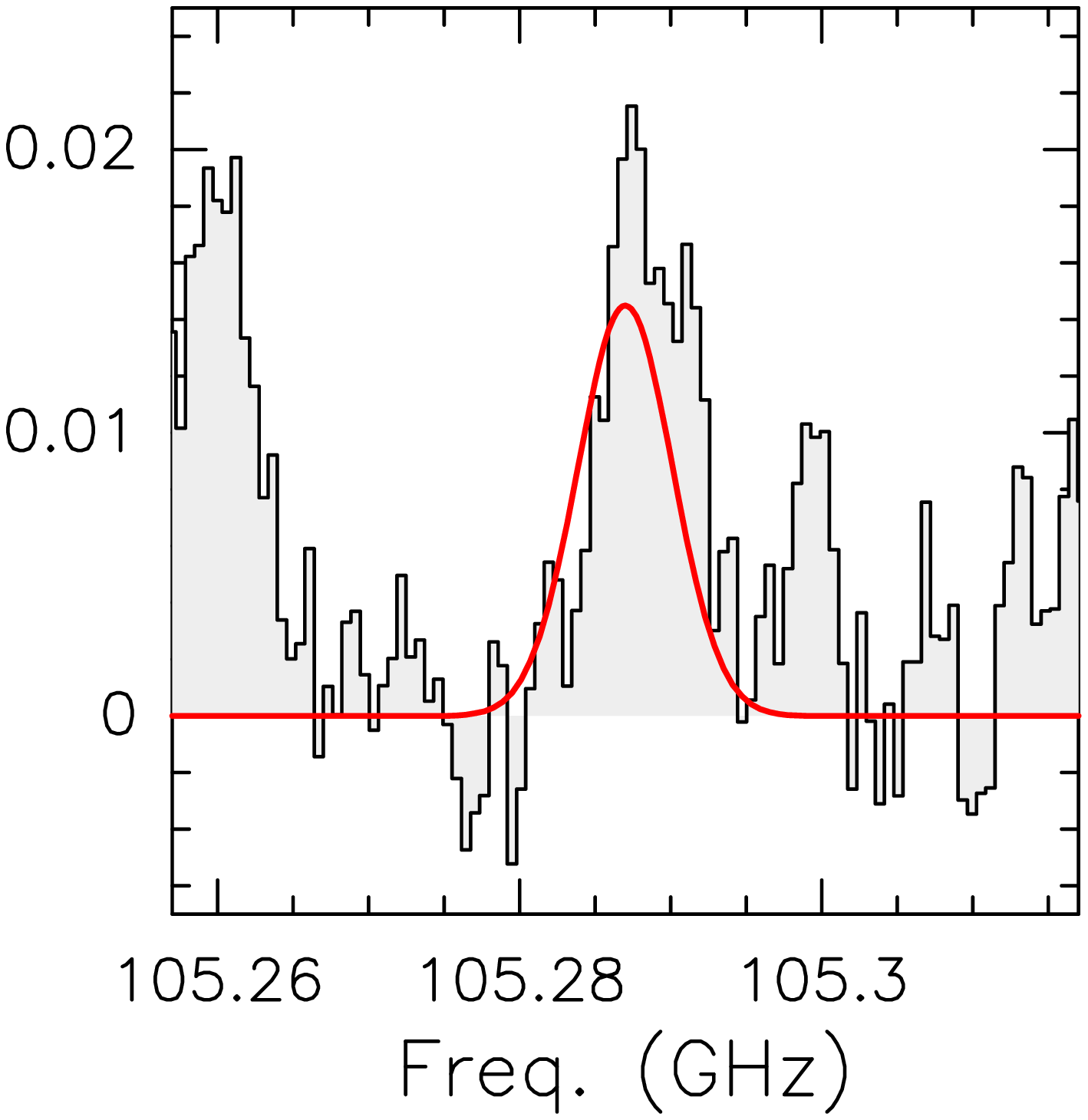}
\centering
\caption{Selected transitions of the syn-conformer of propanal (s-C$_2$H$_5$CHO) detected towards the G+0.693 molecular cloud (see Table \ref{tab:propanal}). The best LTE fit (see Appendix \ref{appendix-c}) is shown in red.}
\label{fig-propanal}
\end{figure*}

The syn-conformer of propanal (s-C$_2$H$_5$CHO) has been identified towards the G+0.693 molecular cloud. 
The analysis has been done using the SLIM-MADCUBA tool, following the same procedure described in Section \ref{sec:analysis}. We use the CDMS entry 58505 (version Jan. 2018). We fixed v$_{\rm LSR}$ to 69 km s$^{-1}$ and $FWHM$ to 21 km s$
^{-1}$, and obtained $T_{\rm ex}$= 12.0$\pm$0.8 K, and $N$=(7.4$\pm$1.5)$\times$10$^{
13}$ cm$^{-2}$, which translates into an abundance with respect to H$_{\rm 2}$ of (5$\pm$2)$\times$10$^{-10}$.
The best LTE fit for selected transitions (listed in Table \ref{tab:propanal}) is shown in Figure \ref{fig-propanal} with a red line.
The s-C$_2$H$_5$CHO transitions that contribute to the spectra shown in Fig. \ref{fig-spectra} are listed in Table \ref{tab:table-contamination}.

\begin{table*}[h]
\centering
\tabcolsep 4.0pt
\caption{Selected transitions of s-C$_2$H$_5$CHO detected towards G+0.693.
We provide the frequencies of the transitions, quantum numbers, the base 10 logarithm of the Einstein coefficients ($A_{\rm ul}$), and the energy of the upper levels ($E_{\rm u}$).}
\begin{tabular}{ c c c c }
\hline
 Frequency & Transition  & log $A_{\rm ul}$  & E$_{\rm u}$   \\
 (GHz) &   &  (s$^{-1}$) & (K)   \\
\hline
77.1819857   & 8$_{1,8}-$7$_{1,7}$ E  & -5.1418  &   17.4   \\  
77.1819857   & 8$_{1,8}-$7$_{1,7}$ A  &  -5.1418 &     17.4   \\ 
77.9521729   & 8$_{0,8}-$7$_{0,7}$ E & -5.1275  &  17.3    \\       
77.9521729   & 8$_{0,8}-$7$_{0,7}$ A & -5.1275   & 17.3    \\ 
 
87.0228499  & 9$_{0,9}-$8$_{0,8}$ E & -4.9808  &   21.5   \\  
87.0228499  & 9$_{0,9}-$8$_{0,8}$ A & -4.9808  &   21.5    \\  

88.5156329  & 8$_{2,6}-$7$_{2,5}$ E & -4.9810  & 21.0     \\
88.5156329  & 8$_{2,6}-$7$_{2,5}$ A & -4.9810  &  21.0    \\
88.5476678  & 3$_{3,1}-$2$_{2,0}$ E & -5.0515  &  8.0    \\
88.5485859  & 3$_{3,1}-$2$_{2,0}$ A   & -5.0116  & 8.0     \\
95.2964249    & 10$_{0,10}-$9$_{1,9}$ E  & -4.8656  &   26.1   \\  
95.2964249      & 10$_{0,10}-$9$_{1,9}$ A & -4.8656  &  26.1    \\ 
95.7920583    & 10$_{1,10}-$9$_{1,9}$ E & -4.8531  & 26.1   \\
95.7920583    & 10$_{1,10}-$9$_{1,9}$ A & -4.8531  & 26.1   \\   
96.1200037      & 10$_{0,10}-$9$_{0,9}$ E & -4.8484  &  26.1  \\   
96.1200037      & 10$_{0,10}-$9$_{0,9}$ A & -4.8484  & 26.1   \\    
99.3945076    & 4$_{3,1}-$3$_{2,2}$ E  & -4.9654  &  10.0  \\   
99.3945076     & 4$_{3,1}-$3$_{2,2}$ A  &  -4.9643  &  10.0  \\   
105.2870083    & 10$_{1,9}-$9$_{1,8}$ E  & -4.7369 & 29.1   \\   
105.2870083     & 10$_{1,9}-$9$_{1,8}$ A  & -4.7369  & 29.1   \\ 
\hline 
\end{tabular}
\label{tab:propanal}
\end{table*}

\clearpage

\section{Detections of HNO and N$_2$O towards G+0.693}
\label{appendix}

\begin{figure*}
\includegraphics[scale=0.5,angle=0]{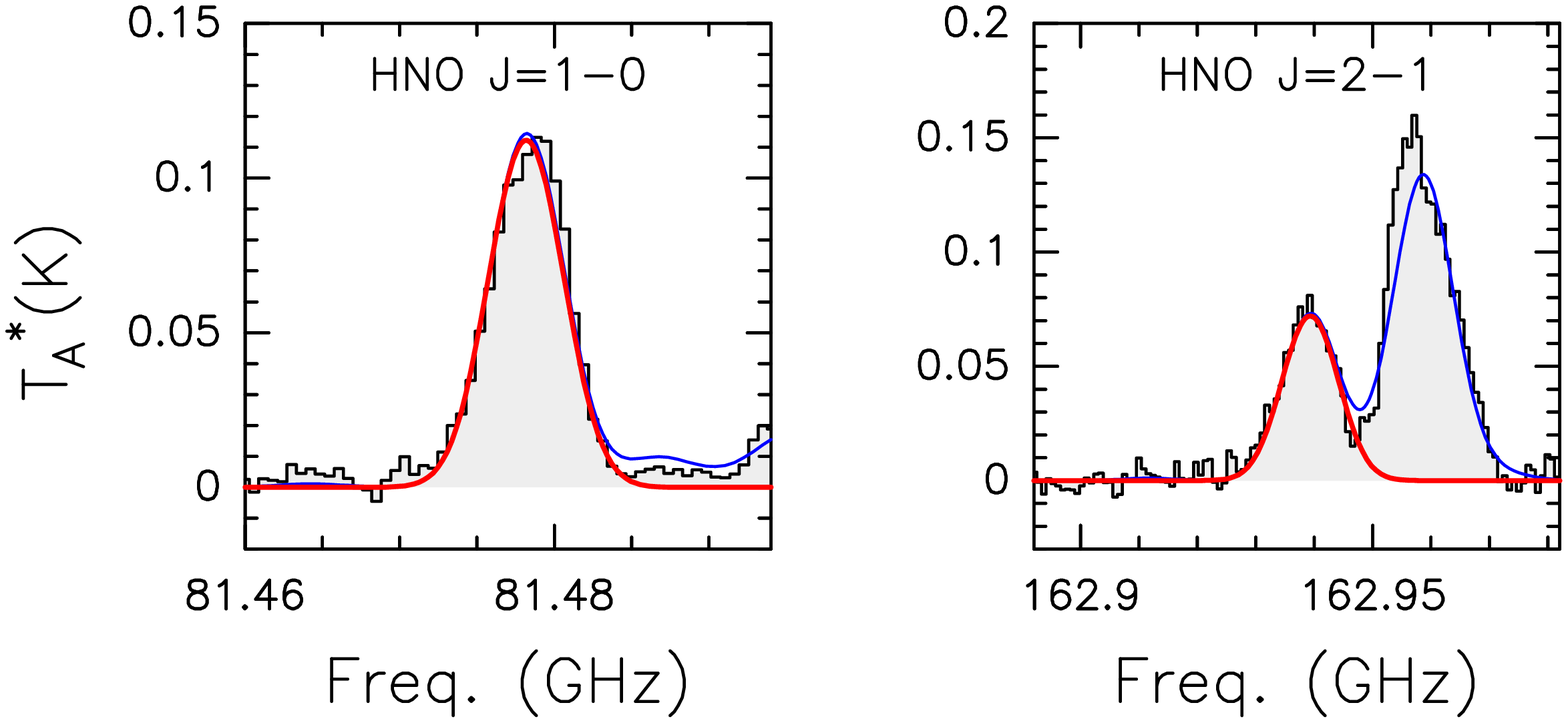}
\centering
\caption{HNO transitions detected towards the G+0.693 molecular cloud. The best LTE fit is shown in red, while the total contribution including also other molecular species identified in our spectral survey is shown in blue.}
\label{fig-HNO}
\end{figure*}

\begin{figure*}
\includegraphics[scale=0.6,,angle=0]{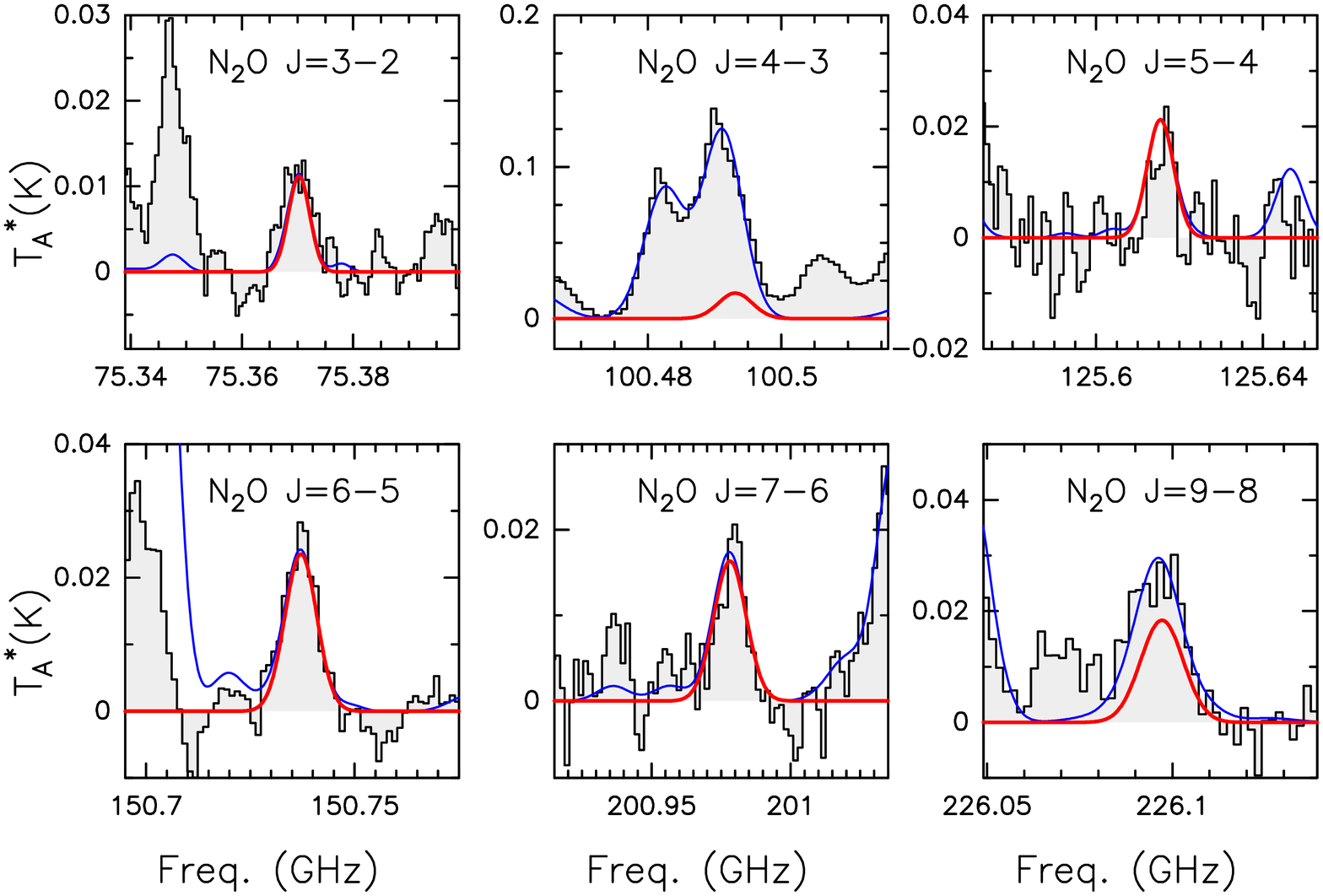}
\centering
\caption{N$_2$O transitions detected towards the G+0.693 molecular cloud. The best LTE fit is shown in red, while the total contribution including also other molecular species identified in our spectral survey is shown in blue.}
\label{fig-N2O}
\end{figure*}

We report here the detection of the nitroxyl radical (HNO) and nitrous oxide (N$_2$O) towards G+0.693.
We show in Figs. \ref{fig-HNO} and \ref{fig-N2O} the detected transitions of HNO and N$_2$O, respectively, which are listed in Table \ref{tab:other-mols}. The analysis have been done using the SLIM-MADCUBA tool, following the same procedure described in Section \ref{sec:analysis}. The derived parameters from the  fits ($N$, $T_{\rm ex}$, $FWHM$ and v$_{\rm LSR}$) are presented in Table \ref{tab:parameters}.

\clearpage
\begin{table*}[h]
\centering
\tabcolsep 4.0pt
\caption{Transitions of HNO (JPL entry 31005, version Feb. 1996) and N$_2$O (JPL entry 44004, version Aug. 2014) detected towards G+0.693.
We provide the frequencies of the transitions, quantum numbers, the base 10 logarithm of the Einstein coefficients ($A_{\rm ul}$), and the energy of the upper levels ($E_{\rm u}$).
}
\begin{tabular}{c c c c c }
\hline
Molecule & Frequency & Transition$^{(a)}$  & log $A_{\rm ul}$  & E$_{\rm u}$   \\
& (GHz) &   &  (s$^{-1}$) & (K)   \\
\hline
HNO & 81.47749   & 1$_{0,1}-$0$_{0,0}$ F=2-1 & -5.65235  &  3.9     \\       
HNO & 81.47749   & 1$_{0,1}-$0$_{0,0}$ F=1-1 & -5.65240  &      3.9 \\  
HNO & 81.47749   & 1$_{0,1}-$0$_{0,0}$ F=0-1 & -5.65240 & 3.9    \\       
HNO & 162.937949   & 2$_{0,2}-$1$_{0,1}$ F=3-2  & -4.67021  & 11.7      \\       
HNO & 162.937949    &  2$_{0,2}-$1$_{0,1}$ F=2-1 & -4.79519  & 11.7      \\       
HNO & 162.937949    &  2$_{0,2}-$1$_{0,1}$ F=1-0 &  -4.92554 & 11.7      \\       
HNO & 162.937949     &  2$_{0,2}-$1$_{0,1}$ F=1-1 & -5.05044  &  11.7    \\       
HNO & 162.937949    &  2$_{0,2}-$1$_{0,1}$ F=2-2 &  -5.27229 & 11.7       \\       
HNO & 162.937949    &  2$_{0,2}-$1$_{0,1}$ F=1-2 & -6.22654  & 11.7      \\ 
\hline
N$_2$O &  75.369224  &  3$-$2     & -7.25771  &   7.2    \\
N$_2$O$^b$ &  100.49174  & 4$-$3  & -6.86711  & 12.1       \\
N$_2$O &  125.61369  & 5$-$4      & -6.56660  &  18.1      \\
N$_2$O &  150.735046  & 6$-$5     & -6.32247  & 25.3       \\
N$_2$O & 200.975333   & 8$-$7     & -6.32247  &  43.4     \\ %
N$_2$O$^c$ & 226.0940229   & 9$-$8    & -6.32247  &  54.3       \\
\hline 
\end{tabular}
\label{tab:other-mols}
{\\ (a): The format for the quantum numbers is $J_{\rm K_a,K_c}$ for HNO and $J$ for N$_2$O.}
{\\ (b): Transition contaminated by CH$_3$OCHO (100.490682 GHz) and aGg-(CH$_2$OH)$_{2}$ (100.4906121 GHz).}
{\\ (c): Transition contaminated by CH$_3$SH (226.093108 GHz).}
\end{table*}

\end{appendices}

\end{document}